\documentclass[12pt,preprint]{aastex}
\usepackage{threeparttable}
\slugcomment{Accepted by ApJ}
\shorttitle{Filament Formation}
\shortauthors{Xia et al.}
\begin{document}

\title{Formation of Solar Filaments by Steady and Nonsteady
	Chromospheric Heating}
\author{C. Xia\altaffilmark{1}, P. F. Chen\altaffilmark{1}, R. Keppens
\altaffilmark{2,3}, A.J. van Marle\altaffilmark{2}}
\altaffiltext{1}{Department of Astronomy, Nanjing University, Nanjing 
210093, China; chenpf@nju.edu.cn}
\altaffiltext{2}{Centre for Plasma Astrophysics, K.U.Leuven, 
Celestijnenlaan 200B, 3001 Heverlee, Belgium}
\altaffiltext{3}{FOM-Institute for Plasma Physics Rijnhuizen, 
Nieuwegein, The Netherlands}

\begin{abstract}
It has been established that cold plasma condensations can form in a
magnetic loop subject to localized heating of the footpoints. In this
paper, we use grid-adaptive numerical simulations of the radiative
hydrodynamic equations to investigate the filament
formation process in a pre-shaped loop with both steady and finite-time
chromospheric heating. Compared to previous works, we consider
low-lying loops with shallow dips, and use a more realistic description
for the radiative losses. We demonstrate for the first time that
the onset of thermal instability satisfies the linear instability
criterion.  The onset time of the condensation is roughly $\sim 2$ hr or
more after the localized heating at the footpoint is effective, and the
growth rate of the thread length varies from 800 km hr$^{-1}$ to 4000 km
hr$^{-1}$, depending on the amplitude and the decay length scale
characterizing this localized chromospheric heating. We show how single
or multiple condensation segments may form in the coronal portion. In
the asymmetric heating case, when two segments form, they approach and
coalesce, and the coalesced condensation later drains down into the
chromosphere. With a steady heating, this process repeats with a
periodicity of several hours. While our parametric survey confirms
and augments earlier findings, we also point out that steady heating is
not necessary to sustain the condensation. Once the condensation is
formed, it keeps growing even after the localized heating ceases. In
such a finite-heating case, the condensation instability is
maintained by chromospheric plasma which gets continuously siphoned into
the filament thread due to the reduced gas pressure in the corona.
Finally, we show that the condensation can survive continuous
buffeting of perturbations from the photospheric {\it p}-mode
waves.
\end{abstract}


\section{Introduction}\label{intro}

Solar filaments are cold and dense plasma concentrations, suspended
magnetically in the hot and tenuous corona, sometimes with barbs
extending from the main spine down to the chromosphere \citep{Prie88,
Tand95}. They appear as dark features in H$\alpha$ on the solar disk,
while they are bright when viewed above the solar limb as prominences.
Typically, filaments appear as a narrow spine above the magnetic
neutral line of the photospheric magnetograms. High-resolution
observations actually revealed that the filament spine is composed of a
collection of separate threads, which are typically 2--20 Mm in length
and 100--200 km in width, which reaches the resolution limit of
modern observations \citep{Engv04, Lin05}.
The individual threads are generally weakly inclined to the 
magnetic neutral line. Both spectral and imaging observations indicate
 that the cold plasma in the threads keeps moving, with mean velocity 
of $\sim$10 km s$^{-1}$ ranging from 5 km s$^{-1}$ to 39 km s$^{-1}$, 
in both directions along the threads \citep{Zirk98, Schm91, Lin03,
Okam07, Berg08, Schm10}. Considering that the plasma beta is low in the 
magnetic surroundings \citep[typically 0.1, see][]{Mack05}, it is 
generally assumed that the motions are channeled by the magnetic field. 
The threads can be considered as the building blocks of filaments, and 
understanding the formation of solar filaments should start with the 
reduced problem of forming a single field-aligned cold thread.

The magnetic configuration supporting filaments can be divided into
two classes \citep{Prie88}, namely, the normal-polarity type
\citep{Kipp57} and the inverse-polarity type \citep{Kupe74}. Both of
them contain a dip above the magnetic neutral line, which is thought to
be important in suspending the heavy filament against gravity.
The existence of magnetic dips was frequently inferred by photospheric
vector magnetograms \citep{Lope06} or found in the extrapolated coronal
force-free field based on the photospheric magnetograms \citep{Aula98,
Yan01, Guo10, Jing10}.

Realizing that an H$\alpha$ thread contains more mass than the coronal
portion of the flux tube, it was suggested that the mass in the threads
originates from the chromosphere \citep{Malh89,
Mack10}. There are basically three types of mechanisms for the
chromospheric mass to fill the coronal portion of a flux tube
\citep[see][for a review]{Mack10}. First, chromospheric plasma can be
injected into coronal loops at the footpoints. The injection may result 
from chromospheric reconnection \citep{Chae01} or from the shearing
motions of the magnetic loop \citep{Choe92}. As the second mechanism,
the chromospheric mass, along with the flux tubes, can be uplifted to
the corona after magnetic cancellation in the chromosphere \citep{vanB90,
Prie96, Litv05}. Another approach involves chromospheric evaporation
from the footpoints of the flux tubes to the coronal portion, which then
triggers a localized coronal condensation. The chromospheric
evaporation, which is also a kind of mass injection, is thought to be
due to localized heating in the low atmosphere \citep{Pola86, Mok90,
Dahl98}.

The last approach was demonstrated numerically by \citet{Anti99},
suggesting that the cold plasma condensation in the filament threads is
due to thermal non-equilibrium or ``catastrophic cooling". It was shown
that as localized heating is introduced in the low atmosphere, chromospheric 
plasma is evaporated into the coronal portion of the flux tube. For a 
uniform heating with an amplitude $\sim$ 10$^{-3}$ erg cm$^{-3}$ s
$^{-1}$, the coronal loop only becomes hot and dense, whereas for a 
localized heating with the same amplitude, the enhanced radiation due 
to optically thin radiative losses in the corona leads to catastrophic 
cooling and plasma condensation. While symmetric heating was assumed in 
these simulations, further simulations showed that steady asymmetric 
heating can yield periodic formations of cold plasma condensations 
across the magnetic dip and their drainage to a footpoint of the flux 
tube \citep{Anti00}. \citet{Karp01} found that even arched field lines 
can also host the repetitive formation and drainage of the cold plasma 
condensation, implying that the magnetic dip might not be a necessary 
condition for the filament formation, though a deeply dipped field 
line hinders the condensation from draining down, keeping the 
H$\alpha$ thread near the magnetic dip for a long time \citep{Karp03}. 
Assuming a more realistic asymmetric loop geometry and a non-uniform 
cross section, together with adopting an updated radiation loss function,
it was 
found that the numerical simulations can reproduce the formation rate, 
the elongated structure of the condensations, and the high speed 
motions ($\sim$ 50 km s$^{-1}$) of the filament thread \citep{Karp05, 
Karp06}. Condensations can also form when the energy input is 
impulsive and randomly distributed in time, provided that the average 
interval between energy pulses is shorter than the coronal radiative 
cooling time \citep[$\sim$2000 s,][]{Karp08}. This thermal 
non-equilibrium model is also used to simulate other condensations in 
coronal loops, such as coronal rains, with a semicircular geometry and 
a shorter length \citep{Mull03, Mull04, Klim10}. All these studies
emphasized that an adaptive mesh is critically necessary to resolve 
the thin transition regions between a condensation and its surrounding
corona and to follow the condensation throughout its evolution.

In most previous works, the dynamic formation of filaments in the
magnetized solar corona is reduced to a one-dimensional (1D) radiative
hydrodynamic problem along a given magnetic loop. The simulation then
tracks the plasma dynamics along the loop under the influence of
gravity, pressure gradients, thermal conduction, optically thin
radiative losses, and a prescribed heating. In all these works, the
strong localized heating was set to be steady or intermittent for tens
of hours. If this localized heating is due to chromospheric 
reconnection, the heating should in reality be short-lasting. It is
still unclear whether a one-off heating with finite lifetime can lead to
the formation of a long filament thread. Starting from the simulations
with steady localized heating (both symmetric and asymmetric), which are
aimed to investigate the details of the plasma condensation and its
dynamics, this paper, for the first time, further investigates the
response of the coronal loop on the localized heating with a limited
duration. In addition, due to the large contemporary interest in 
prominence seismology, we address whether quiescent prominences (or 
threads) can survive the continuous perturbations from the $p$-mode 
waves, and how these wave modes get transmitted and reflected through 
the prominence body. The paper is organized as follows. Our numerical 
method is described in \S\ref{method}, and the results for steady 
heating, which confirm and extend earlier work to a wider parameter 
regime are presented in \S\ref{result}. Section \ref{discuss} collects 
all novel aspects of our work: (1) confronting the evolution with the 
criteria of the thermal instability, which accounts for the catastrophic 
cooling; (2) the response of the plasma condensation on the switch-off 
of the localized heating; and (3) the stability of the condensation 
under {\it p}-mode driven perturbations. Conclusions are drawn in 
\S\ref{con}.

\section{Numerical Method}\label{method}

\subsection{Governing equations and radiative loss treatment}

\begin{figure}
\includegraphics[width=6.in]{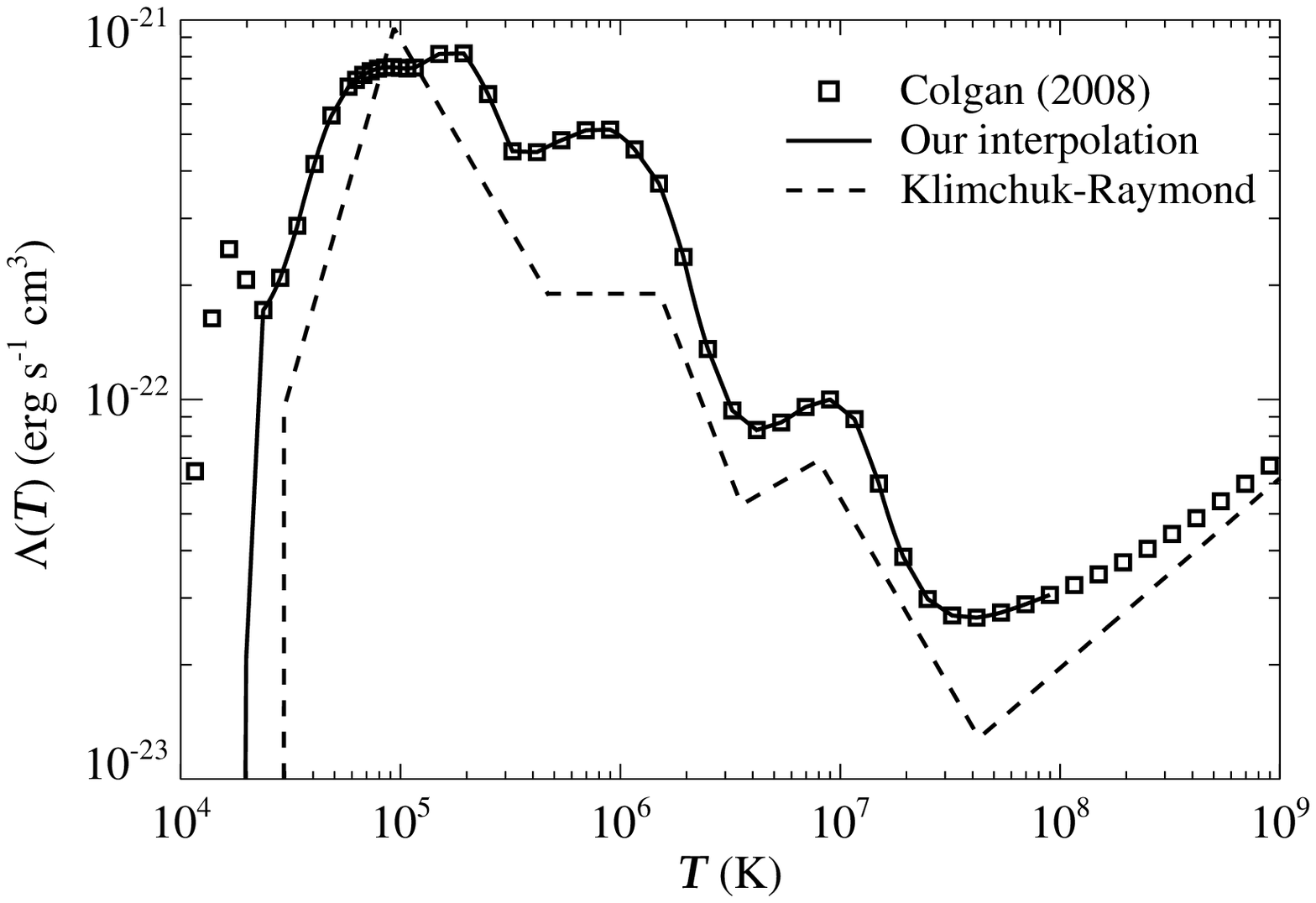}
\caption{The radiative loss coefficient $\Lambda(T)$ vs. $T$ calculated
by \citet{Colg08} ({\it squares}) and our interpolation ({\it solid
line}). Note that the Klimchuk-Raymond profile ({\it dashed line}) shows 
the piece-wise continuous radiative loss function used in previous works.}
\label{fig:cooling}
\end{figure}

As mentioned above, the plasma beta of the filament environment is
believed to be small, therefore it is generally assumed that the coronal
flux tubes, which can support the filaments against gravity are rigid
and the mass flow is channeled along the magnetic field line in the
corona \footnote{It is noted that the plasma condensation greatly
enhances the effect of the gravity, which may deform the magnetic loop
as demonstrated by \citet{Wu90}.}. With such an assumption, the plasma 
dynamics of the filament threads is simply described by the 
1D radiative hydrodynamic equations as follows:

\begin{equation}\label{eq1}
\frac{\partial \rho}{\partial t} + \frac{\partial}{\partial s} (\rho v)
	= 0,
\end{equation}

\begin{equation}\label{eq2}
\frac{\partial}{\partial t} (\rho v) + \frac{\partial}{\partial s}
	(\rho v^2 + p) = \rho g_\parallel (s),
\end{equation}

\begin{equation}\label{eq3}
\frac{\partial \varepsilon}{\partial t} + \frac{\partial}{\partial s}
	(\varepsilon v+ p v)=\rho g_\parallel v+H(s)-n_{\rm H} n_{\rm e} 
   \Lambda(T)+\frac{\partial}{\partial s}\left(\kappa \frac{\partial T}
	{\partial s}\right),
\end{equation}

\noindent
where $\rho$ is the mass density, $T$ is the temperature, $s$ is the
distance along the loop, $v$ is the velocity of plasma, $p$ is the gas
pressure, $\varepsilon=\rho v^2 /2 + p/(\gamma -1)$ is the total energy
density, $n_{\rm H}$ is the number density of hydrogen, $n_{\rm e}$ is 
the number density of electrons, and $g_\parallel(s)$ is the component
of gravity at a distance $s$ along the magnetic loop.  Furthermore, 
$\gamma=5/3$ is the ratio of the specific heats, $\Lambda(T)$ is the 
radiative loss coefficient for the optically thin emission, $H(s)$ is
the volumetric heating rate, and $\kappa=10^{-6} T^{5/2}$ erg cm$^{-1}$ 
s$^{-1}$ K$^{-1}$ is the Spitzer heat conductivity. As done in 
previous works mentioned in \S\ref{intro}, we assume a fully ionized 
plasma and adopt the one-fluid model. Considering the helium abundance
($n_{\rm He}/n_{\rm H}=0.1$), we take $\rho=1.4 m_p n_{\rm H}$ and
$p=2.3 n_{\rm H} k_B T$, where $m_p$ is the proton mass and $k_B$ is
the Boltzmann constant. The radiative hydrodynamic equations (\ref{eq1}
--\ref{eq3}) are numerically solved by the Adaptive Mesh Refinement 
Versatile Advection Code (AMRVAC) \citep{Kepp03,Kepp11}, where the heat
conduction term is solved with an implicit scheme separately from other
terms. To calculate the radiative energy loss, we use a second order
polynomial interpolation to compile a high resolution table based on the
radiative loss calculations recently done by \citet{Colg08}. They
calculated the radiative losses for the solar coronal plasma using a
recommended set of quiet region element abundances. In their 
calculations they used a complete and self-consistent atomic data set
and an accurate atomic collisional rate over a wide temperature range.
As shown in Figure \ref{fig:cooling}, $\Lambda(T)$ in our cooling table
({\it solid line}) interpolated from \citet{Colg08} ({\it square}) is
generally $\sim 2$ times larger than the Klimchuk-Raymond radiative loss
function ({\it dashed line}) used in previous works \citep{Karp05, 
Karp06, Karp08, Klim10}. The figure also demonstrates that our 
cooling curve better represents the detailed temperature dependence of 
the radiative loss.

Using our cooling table, we then exploit the exact integration
scheme \citep{Town09}, rather than traditional implicit or explicit time
stepping methods. This method is much faster than an explicit scheme, as
it can avoid the numerical limit of the radiative timescale on the
simulation timestep. Besides, it is more stable than the implicit
schemes based on Newton-Raphson iterations. Below 20000 K, we set
$\Lambda(T)$ to vanish since the plasma then becomes optically thick and
is no longer fully ionized. The use of explicit, (semi-)implicit and
exact integration methods in grid-adaptive simulations was analyzed
recently by \citet{vanM11}.

\subsection{Discretization and AMR settings}

When using the AMRVAC code, the Total Variation Diminishing 
Lax-Friedrichs (TVDLF) scheme using linear reconstruction employing 
a Monotonized Central limiter \citep{toth96}, is chosen for the spatial 
differentiation, combined with a predictor-corrector two-step explicit 
scheme for the time progressing. Six levels of adaptive mesh refinement 
(AMR) in a block-based AMR approach are applied, which leads to a 
minimum grid spacing of 6.77 km, comparable to 5--6 km in previous 
works such as \citet{Klim10}. The refinement/coarsening criteria are 
based on numerical errors estimated using density and its gradient 
following Lohner's prescription \citep{lohner}. If any error exceeds 
0.1, the block is refined. If all errors in the block are less than 
0.0125, the block is then coarsened. To include the heat conduction 
source in the energy equation, we separately solve the heat conduction 
term in each AMR grid block, using the implicit scheme where the 
central difference is taken for the space derivative of the 
temperature. This leads to a local tri-diagonal linear system per grid 
block, where the temperature in the block boundaries is taken from 
neighboring blocks at previous time step. To simulate 10 hr physical 
time, our implementation uses $\sim 1.5$ hr on 4 processors.

\subsection{Initial and boundary conditions}

\begin{figure}
\includegraphics[width=6.in]{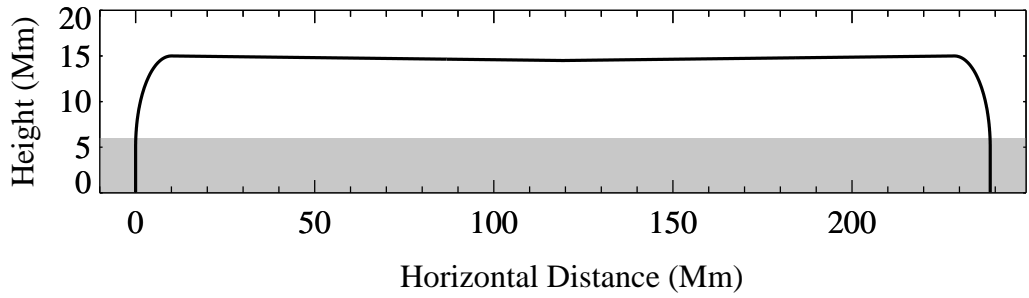}
\caption{Geometry of our model loop, which represents a magnetic field
line across a filament thread. The gray rectangular region denotes the
photosphere and the chromosphere. Note that the vertical and the
horizontal axes are not to scale.}
\label{fig:loop}
\end{figure}

We adopt a loop geometry with a magnetic dip, which is symmetric about
the midpoint. On each side, the loop has a vertical leg of 5 Mm in
length above the footpoint and a quarter-circular arc of 15.7 Mm in
length connecting the vertical leg and the dip, which is 218.6 Mm in
length, as shown in Figure \ref{fig:loop}. Note that the geometry of the
loop determines the distribution of $g_\parallel(s)$, which is symmetric 
about the midpoint and whose left half is described as follows:

\begin{equation}
g_\parallel(s)=\cases{-g_\odot , & $s \leqslant s_1$; \cr
-g_\odot \cos\left(\frac{\displaystyle \pi}{\displaystyle 2}\frac
{\displaystyle s-s_1}{\displaystyle s_2-s_1}\right), & 
$s_1 < s \leqslant s_2$; \cr
g_\odot \frac{\displaystyle \pi D}{\displaystyle 2(L/2-s_2)} 
\sin\left(\pi \frac{\displaystyle s-s_2}
{\displaystyle L/2-s_2}\right), & $s_2 < s \leqslant L/2$, \cr}
\end{equation}
where $g_\odot=2.7 \times 10^4$ cm s$^{-2}$ is the solar gravity,  
$s_1=5$ Mm, $s_2=s_1+15.7$ Mm, $L=260$ Mm is the total loop length, 
and $D=0.5$ Mm is the dip depth. The value of total loop length $L$ 
is suggested by the observations of \citet{Okam07}. The dip is very 
shallow, so the coronal part of the loop is nearly flat. The 
midpoint of the loop, which is the center of the dip, has a height 
of 14.5 Mm above the bottom boundary.

\begin{figure}
\includegraphics[width=6.in]{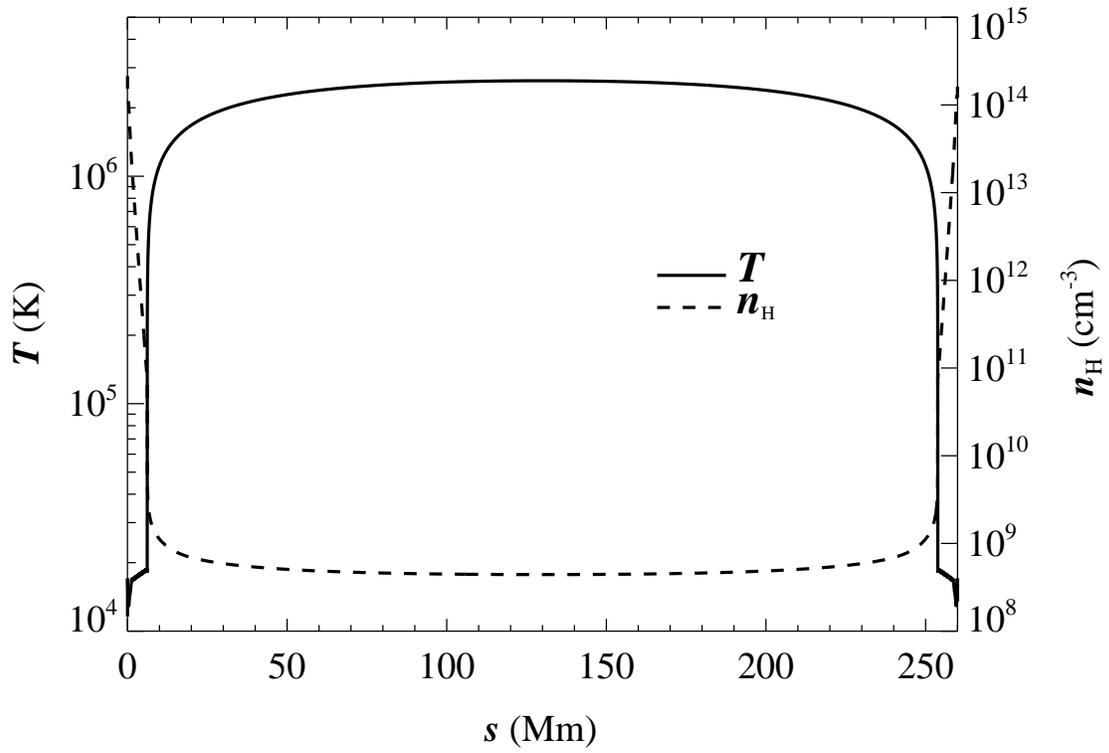}
\caption{Distributions of the temperature ({\it solid line}) and the
number density of hydrogen ({\it dashed line}) along the model loop in 
the initial hydrostatic state.}
\label{fig:init}
\end{figure}

The initial equilibrium state is obtained by numerically solving
Equations (\ref{eq1}--\ref{eq3}). We start with a temperature
($T$) versus height ($h$) distribution of $T=\tanh (h-h_0)$ with 
$T=10^6$ K in the corona and 6000 K in the photosphere, which is close 
to the quiet Sun atmospheric model \citep{Vern81}. The density is 
determined by balancing the pressure gradient with the gravity, with 
$n_H=\rho/(1.4 m_p)=10^9$ cm$^{-3}$ at the loop center. At this stage, 
only the background heating $H_0(s)$ is included in the energy 
equation, namely $H(s)=H_0(s)$ in Equation (\ref{eq3}). $H_0(s)$ is a 
steady term in order to maintain the hot corona, whose physics is 
still under debate. Considering that the photospheric motions are the 
source of the energy that is transported upward to heat the 
chromosphere and the corona somehow, it is generally conjectured that 
the heating rate decays with height \citep{Seri81, Mok90, Asch02}. 
Similar to previous works, we assume that $H_0(s)$ exponentially 
decreases with the distance away from the nearest footpoint along the 
loop, and remains constant in time:

\begin{equation}
H_0(s)=\cases{E_0 \exp(-s/H_m), & $s < L/2$; \cr
        E_0 \exp[-(L-s)/H_m], & $L/2 \leqslant s < L$, \cr}
\end{equation}
\noindent
where the amplitude $E_0= 3 \times 10^{-4}$ ergs cm$^{-3}$ s$^{-1}$
\citep{With77}, and the scale length $H_m=L/2$ \citep{With88}. The 
prescribed distributions are in force equilibrium, but not in thermal 
equilibrium, and will evolve to reach a new hydrostatic state. Such a 
state, whose density and temperature distributions are displayed in 
Figure~\ref{fig:init}, serves as the initial conditions for our further 
simulations. In the initial state, the temperature is the highest at 
the midpoint of the loop, with $T=2.6\times 10^6$ K and $n_H=3.2\times 
10^8$ cm$^{-3}$. The thin transition layer between the low atmosphere 
and the corona is roughly at a height of $s_{tr}=6$ Mm. In the low 
atmosphere, $T$ ranges from 13000 K to 18000 K, which is closer 
to the quiet Sun atmospheric model \citep{Vern81} than previous works 
\citep{Karp01,Karp06}. The number density at the endpoints is about 
$2\times 10^{14}$ cm~$^{-3}$. The simulated chromosphere and 
photosphere are about twice thicker than the real Sun, and serve as a 
mass reservoir for the chromospheric evaporation.

For the boundary conditions, we fix the density, velocity, and
temperature at the two endpoints of the loop. Because the 
density in the photosphere is more than 4 orders of magnitude higher than
those in the filaments and the corona, the coronal dynamics has little
effect on the photosphere, justifying these fixed boundary conditions.

Similar to \citet{Anti99}, in order to simulate the chromospheric
evaporation, an extra localized heating $H_l(s)$, which might be due to
chromospheric reconnection, is added to the energy equation in addition
to the background heating $H_0(s)$, namely $H(s)=H_0(s)+H_l(s)$ in
Equation (\ref{eq3}). As described as follows, $H_l(s)$ is uniform in
the photosphere and chromosphere, and decays exponentially with the 
distance away from the nearest chromosphere along the loop with a 
scale length $\lambda$:

\begin{equation}
H_l(s)=\cases{E_1, & $s \leqslant s_{tr}$; \cr
     E_1 \exp[-(s-s_{tr})/\lambda], & $s_{tr} < s \leqslant L/2$; \cr
   f E_1 \exp[-(L-s_{tr}-s)/\lambda], & $L/2 < s \leqslant L-s_{tr}$; \cr
   f E_1, & $s > L-s_{tr}$, \cr}
\end{equation}
where the amplitude $E_1=10^{-2}$ergs cm$^{-3}$ s$^{-1}$ \citep[cf.][]
{With77, Asch01}, $s_{tr}=6$ Mm is the height of transition region, and
the factor $f$ is the ratio of the localized heating rate near the right
footpoint to that near the left. The localized heating $H_l(s)$ is
ramped up linearly over 1000 s, and maintained since then. In this
paper, we numerically investigate two situations, with symmetric and
asymmetric heating, respectively. The parameters in several typical
cases are listed in Table \ref{tab1}. In the symmetric case, a parameter
survey is performed, including the effects of $\lambda$ and $E_1$.

To better identify in which way our simulations augment the
knowledge gained in prominence formation over the last decade,
we list the most important parameters in similar works on radiative
condensation due to localized heating in Tables~\ref{tab2}-\ref{tab3}.
These tables show that our work differs in a variety of aspects,
connected to the overall loop geometry, to the spatio-temporal
prescription of the heating applied, and also notably in the cooling
table used to quantify radiative losses. Motivated by observations
of active region prominences by~\citet{Okam07}, our model loop shape
represents a low-lying, shallowly-dipped loop with a more realistic
scale for its vertical legs and chromospheric height region. In this
shallow dip configuration, our parametric survey explores a wide
range in the heating parameters.

\section{Numerical Results}\label{result}

\subsection{Symmetric Evolution}\label{sym}

\begin{figure}
\includegraphics[width=6.in]{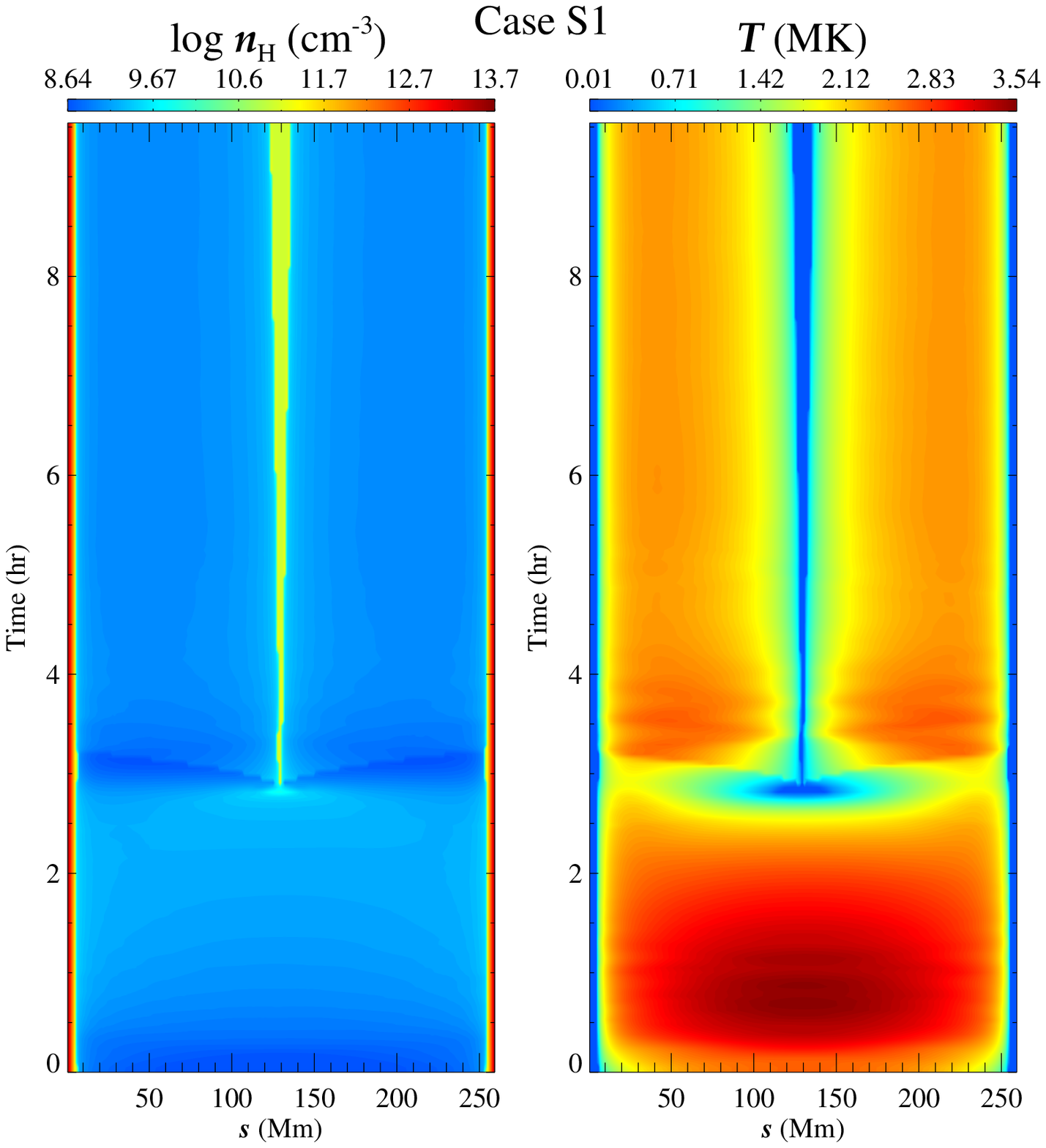}
\caption{Temporal evolution of the number density of hydrogen ({\it left})
and the temperature ({\it right}) along the model loop in case S1. The
two loop footpoints are at $s=0$ and $260$ Mm, respectively, and the 
center of the loop dip is at $s=130$ Mm.}
\label{fig:S1}
\end{figure}

\begin{figure}
\includegraphics[width=6.in]{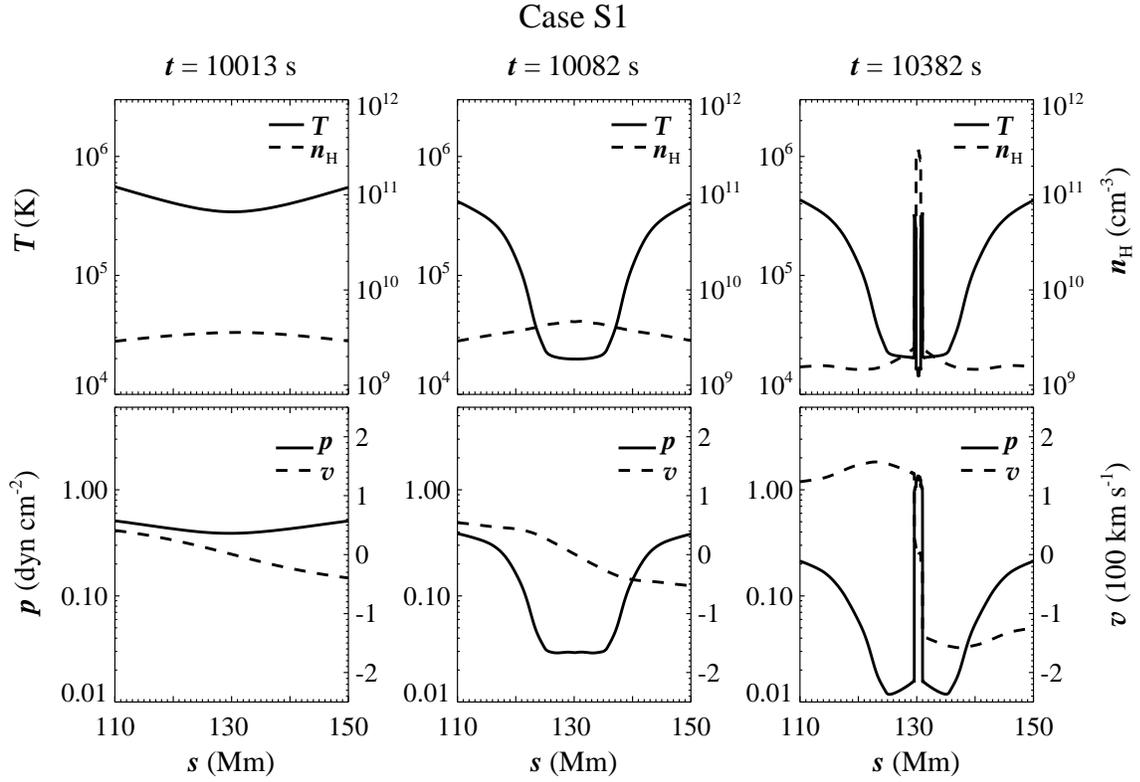}
\caption{Distributions of the temperature ({\it solid line, top panel}), 
the density ({\it dashed line, top panel}), the pressure ({\it solid 
line, bottom panel}), and the velocity ({\it dashed line, bottom panel}) 
across the loop center in case S1 at three moments, i.e., $t=10013$ s 
({\it left column}), $t=10082$ s ({\it middle column}) and $t=10382$ s 
({\it right column}).}
\label{fig:conden}
\end{figure}

As the symmetric localized heating with $\lambda=10$ Mm and $f=1$ is
introduced in case S1, the chromospheric plasma is heated and evaporated
into the corona. As illustrated by Figure \ref{fig:S1}, both the density
$\rho$ and temperature $T$ in the coronal portion increase accordingly.
Near the midpoint of the loop, $T$ reaches the maximum value, $3.49
\times 10^6$ K, at $t=2664$ s, then starts to decline slowly, whereas
$\rho$ keeps increasing slowly from the beginning. At $t=10013$ s, the
temperature and the pressure near the midpoint begin to collapse
simultaneously, drastically decreasing by nearly one and a half orders
of magnitude within 1 min, creating a low pressure cold region, which
expands to a maximum length of 28.4 Mm. At this stage, the density is 
still low, increasing gently as seen from the left and the middle
columns of Figure \ref{fig:conden}. 
Since only the density is chosen to automatically refine the mesh and 
at this stage the density is still smooth, the grid resolution is 108 
km per cell. Even though, this cold region contains 263 grid cells, 
which are sufficient to resolve the region. Due to the large pressure 
gradient that forms at the edge of this cold region, the coronal 
plasma outside the cold region is driven to move rapidly towards this 
central cold region. Since time $t=10082$ s, the density inside this 
cold well begins to increase rapidly as the inflows from two sides 
converge towards the midpoint and compress the cold region. The 
converging velocity reaches 185 km s$^{-1}$. These inflows are 
supersonic with a local Mach number up to 7.

\begin{figure}
\includegraphics[width=6.in]{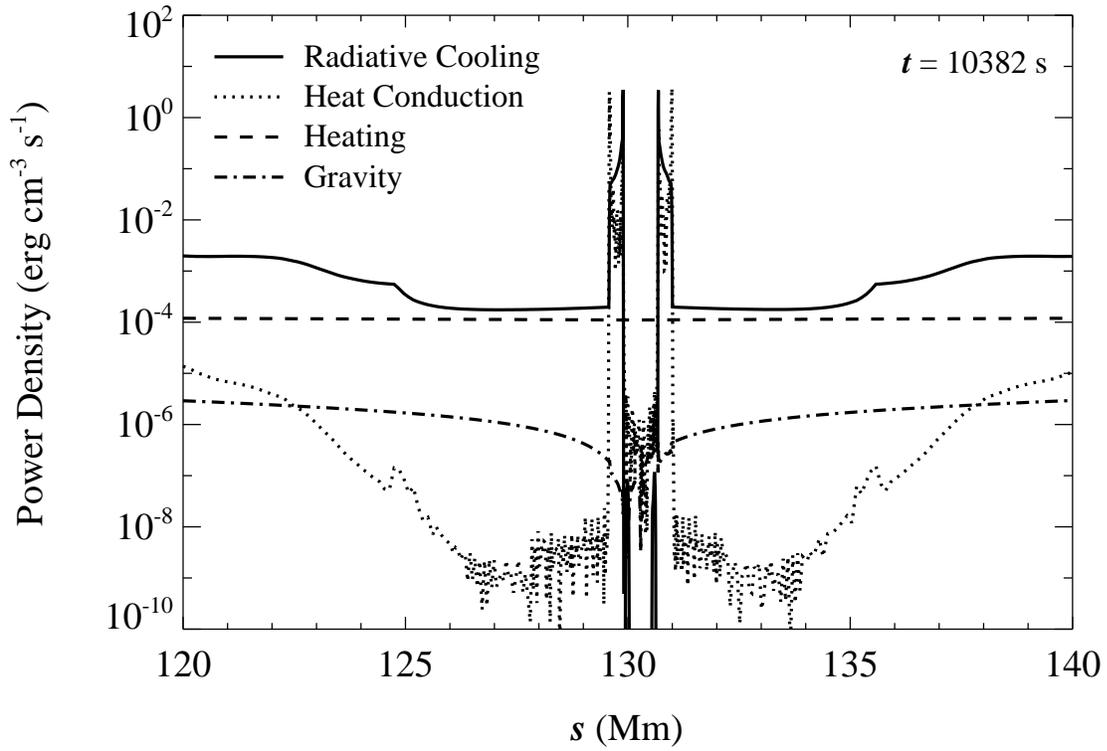}
\caption{The absolute value distributions of various energy sources, 
including the radiative cooling ({\it solid line}), the heat conduction 
({\it dotted line}), the heating ({\it dashed line}), and the gravity 
({\it dashed dotted line}), near the midpoint of the loop at $t=10382$ 
s in case S1.}
\label{fig:energy}
\end{figure}

\begin{figure}
\includegraphics[width=6.in]{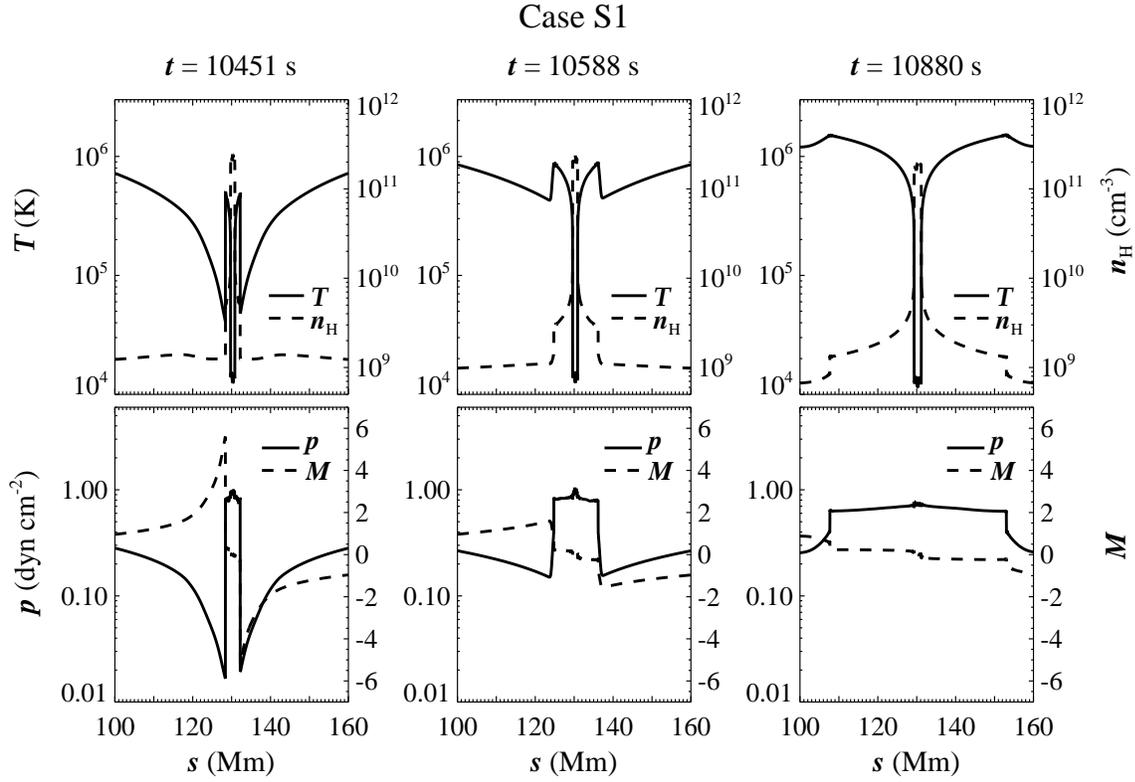}
\caption{Distributions of the temperature ({\it solid line, top}), the 
density ({\it dashed line, top}), the pressure ({\it solid line, bottom}), 
and the Mach number ({\it dashed line, bottom}) along the loop near the 
midpoint, at three instants after those shown in Figure~\ref{fig:conden}, 
namely $t=10451$ s ({\it left column}), $t=10588$ s ({\it middle column}) 
and $t=10880$ s ({\it right column}), in case S1.}
\label{fig:shock}
\end{figure}

As a result, the inflows collide at the midpoint of the loop where a
high pressure peak appears, exciting two rebound shock waves launched 
from the midpoint towards the two sides. A small cold condensation
region ($\sim 1$ Mm in length) is left behind the shocks near the
midpoint at $t\sim 10382$ s (see the right column of Figure
\ref{fig:conden}). To see the contributions of the various terms in
the energy equation, in Figure \ref{fig:energy} we plot the absolute 
value distribution of the energy source terms, including the radiative
cooling, the heat conduction, the heating, and the gravity potential,
across the magnetic dip at $t=10382$ s, when the condensation happens.
It is found that the radiative cooling dominates in the coronal parts
and the boundaries of the condensation segment, but nearly vanishes
inside the condensation. The heat conduction is less important than the
heating in most regions except at the boundaries of the condensation.
The gravity potential is always negligible. The pressure in the cold
region recovers due to the compression of the inflows from outside.
Swept by the outward-propagating rebound shock waves, the depressed
pressure outside the cold region also recovers, as illustrated by Figure
\ref{fig:shock} showing similar quantities at times later than those of
Figure \ref{fig:conden}. The shock waves are bounced back and forth for
$\sim 3$ times between the loop footpoint and the loop center, as
revealed by the sinusoidal pattern in the right panel of Figure 
\ref{fig:S1} between $t=3$ hr and $t=4$ hr. During their passage, they
dissipate their energy to compress and heat the local plasma. The
damping rate is enhanced by thermal conduction and radiation. The plasma
condensation remains near the midpoint, with a temperature of $1.8 
\times 10^4$ K and a density of $1.2\times 10^{11}$ cm$^{-3}$. As a 
contrast, the corresponding values in the neighboring corona are $2 
\times 10^6$ K and $1.03 \times 10^9$ cm$^{-3}$, respectively. Note 
that the condensation temperature is just below 20000 K, where the 
radiative loss is set to vanish smoothly. Further tests indicate that 
if the radiative losses vanish below a lower temperature, the 
condensation would be cooler accordingly. 

\begin{figure}
\includegraphics[width=6.in]{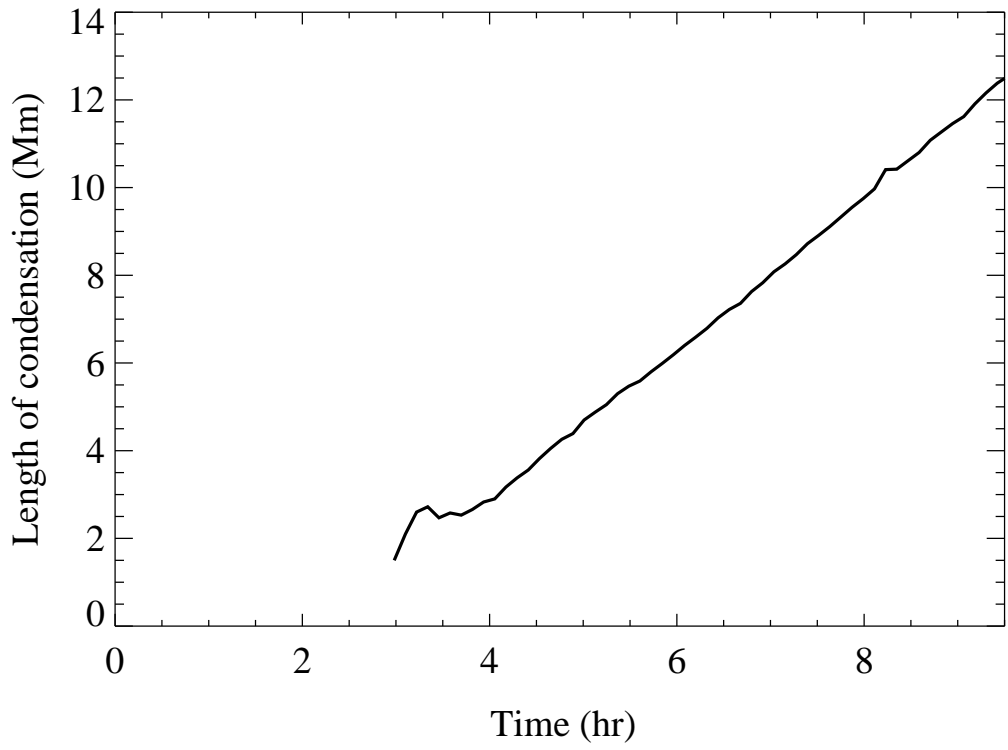}
\caption{Temporal evolution of the length of the condensation in case
S1. The starting point of the line marks the onset of the plasma
condensation.}
\label{fig:grow}
\end{figure}

Figure \ref{fig:grow} depicts the growth of the condensation segment
(or the filament thread). It is seen that the onset time of the
condensation is at $t=2.8$ hr after the localized heating is
introduced in case S1. The growing process of the condensation can be
described as follows: Its length increases rapidly for $\sim 20$ min as
the onset of condensation drives fast evaporation flows from the
chromosphere, which is followed by a slight shrinkage of the
condensation for $\sim 10$ min. As the evaporated plasma flow becomes
steady, the condensation length increases linearly with time, with a
growth rate of 1511 km hr$^{-1}$. With such a speed, it would take 
$\sim 6.6$ hr to form a filament thread with a typical length of 10 Mm.
Observations show that active region filaments form within a day
\citep{Wang07}. For comparison, in \citet{Anti99}, it takes 8.3 hr for a
condensation to grow to 10 Mm long. One reason is that they used a
deeply dipped magnetic loop, where the gravity scale height was shorter,
so that the condensation was strongly squeezed. 

\begin{figure}
\includegraphics[width=6.in]{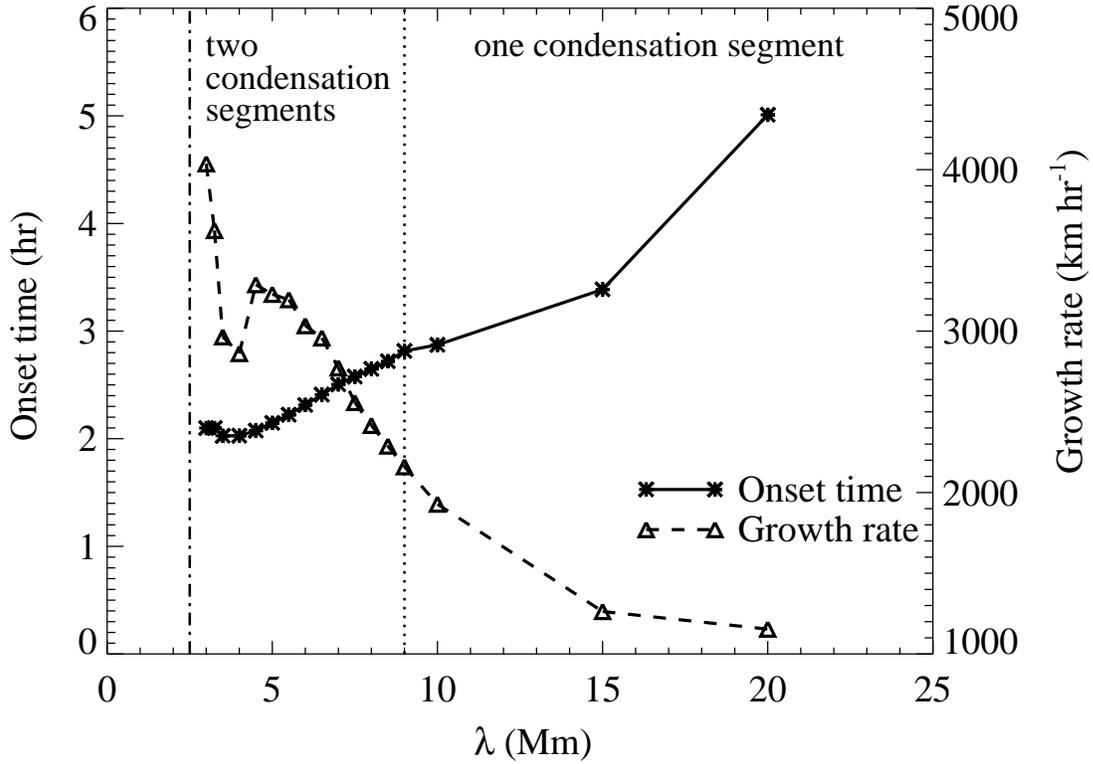}
\caption{Dependence of the onset time ({\it connected asterisks}) and
the mean growth rate ({\it connected triangles}) of the plasma
condensation on the scale height of the localized heating, $\lambda$. 
Note that the figure is divided into two regions where two condensation 
segments or one condensation segment forms by vertical lines.}
\label{fig:lam}
\end{figure}

To investigate the effect of the heating scale length $\lambda$, we
change its value and perform a series of simulations in 17 runs, with
other parameters the same as in case S1. As seen from Figure
\ref{fig:lam}, the onset time of the condensation roughly increases with
$\lambda$, with a minimum of 2 hr. However, the growth rate decreases
with increasing $\lambda$, except a drop down near $\lambda=4$ Mm. It is
noted that if $\lambda$ is larger than 9 Mm, i.e., 1/28 of the total
loop length $L$, the evolution is similar to case S1 (where $\lambda=10$
Mm) as described above, where only a single condensation forms near the
midpoint of the loop. When 3 Mm $<\lambda <$ 9 Mm, i.e., $1/86<\lambda
/L <1/28$, two condensation segments would form first on the two
shoulders of the magnetic dip symmetrically about the midpoint. The two
segments move convergently towards the midpoint, during which both $T$
and $p$ in the region between the two segments drop down. Under this
pressure gradient, the two condensation segments are accelerated from
$\sim 12$ km s$^{-1}$ to 75 km s$^{-1}$, to finally coalesce near the
midpoint. Similar high-speed motion is discussed by \citet{Karp06}.
As $\lambda$ decreases, the two segments form further away
from each other and from the midpoint of the loop. When 2.5 Mm 
$<\lambda <$ 3 Mm, the two condensation segments form in the loop legs 
and then drain down rapidly to the nearby footpoints. When $\lambda <
2.5$ Mm, i.e., $\lambda/L <1/100$, no condensation forms, and the loop
relaxes to a hydrostatic state in the end. Similar situations happen
when $\lambda > 25$ Mm, i.e., $\lambda/L >1/10$, which is the same 
result mentioned by \citet{Klim10}.

\citet{Mull04} simulated the formation of condensations in a 
semicircular coronal loop with a length of 100 Mm and the loop top
temperature of $6.8 \times 10^5$ K. Their loops are shorter and cooler
than our dipped loops. They found that when $\lambda/L = 1/20$, only one
condensation forms at the midpoint of the loop and that two condensation
segments form at shoulders of the loop when $\lambda/L = 1/33$ or 
$\lambda/L = 1/50$. The transition between one condensation and two
condensation segments is somewhere between $\lambda/L =1/20$ and
$1/33$, which is consistent with our result, i.e., $1/28$.
The transition can be understood as follows: At the beginning, the 
plasma at the two shoulders of the loop is cooler and denser than 
that at the midpoint, which means the radiative loss is stronger at
two shoulders. Meanwhile, if $\lambda$ is long enough, the heating 
at shoulders is strong enough to slow down the cooling making the 
midpoint to be the fastest cooling place, and only one condensation forms.
When $\lambda$ decreases, the heating at shoulders is reduced and become 
too weak to obstruct the fast cooling there. Hence,two cold segments are 
formed at the two shoulders before merging near the midpoint of the loop.

\begin{figure}
\includegraphics[width=6.in]{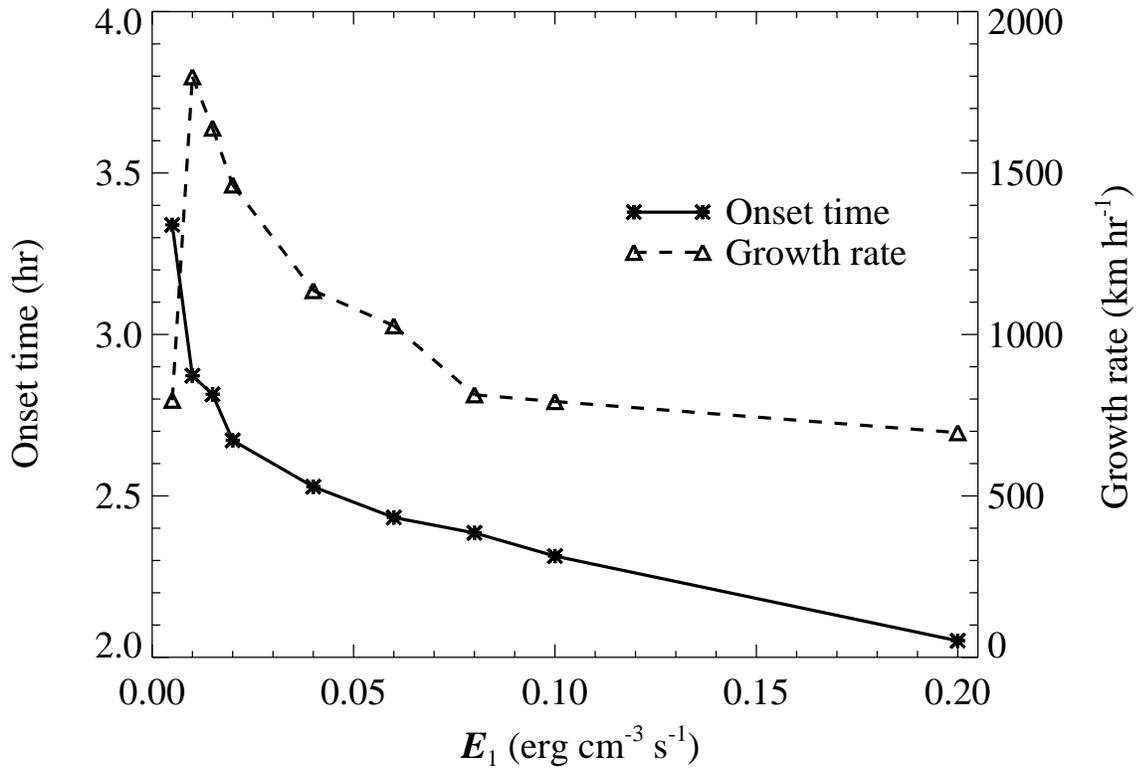}
\caption{Dependence of the onset time of the condensation ({\it
connected asterisks}) and the mean growth rate ({\it connected
triangles}) on the amplitude of the localized heating, $E_1$.}
\label{fig:E1}
\end{figure}

Keeping $\lambda=10$ Mm, we perform another series of simulations with
different amplitude of the localized heating, $E_1$, i.e.,
0.005--0.2 ergs cm$^{-3}$ s$^{-1}$. So the localized heating still
dominates compared to the weak background heating. The evolution is
similar to case S1, such that only one condensation forms near the
midpoint. As indicated by Figure \ref{fig:E1}, the onset time of the
condensation decreases as $E_1$ increases, whereas the growth rate of
the condensation is maximal at $E_1\sim 0.01$ ergs cm$^{-3}$ s$^{-1}$. 
It is easy to understand that the condensation forms earlier with a 
larger $E_1$ since a stronger chromospheric heating leads to stronger
evaporation. The growth rate might be determined by the compromise
between the evaporation rate and the deposited energy. A stronger
heating drives stronger chromospheric evaporation on one hand, and
refrains the evaporated plasma from cooling on the other hand. Another
important factor is that the plasma density in the condensation segment
increases due to higher compression as $E_1$ increases. It becomes
slower for a denser condensation to grow. As a combined result, the
growth of the condensation peaks at $E_1\sim 0.01$ ergs cm$^{-3}$
s$^{-1}$, and decreases at larger $E_1$.

Comparing the onset time and growth rates deduced from observations 
may help to further pin down the properties of the employed localized 
heating. However, to determine the onset time of the thread formation 
requires combined spectral and imaging observations with high 
resolution, which are not available yet. The growth speed of the 
filament thread can be compared with future observations.

\subsection{Asymmetric Evolution}\label{asym}

It is more general that the localized heating in the chromosphere is 
not symmetric between the two footpoints of a magnetic loop. In this 
subsection, we perform three simulations with different $f$, the ratio
of the heating rate at the right footpoint to the left. 

\begin{figure}
\includegraphics[width=6.in]{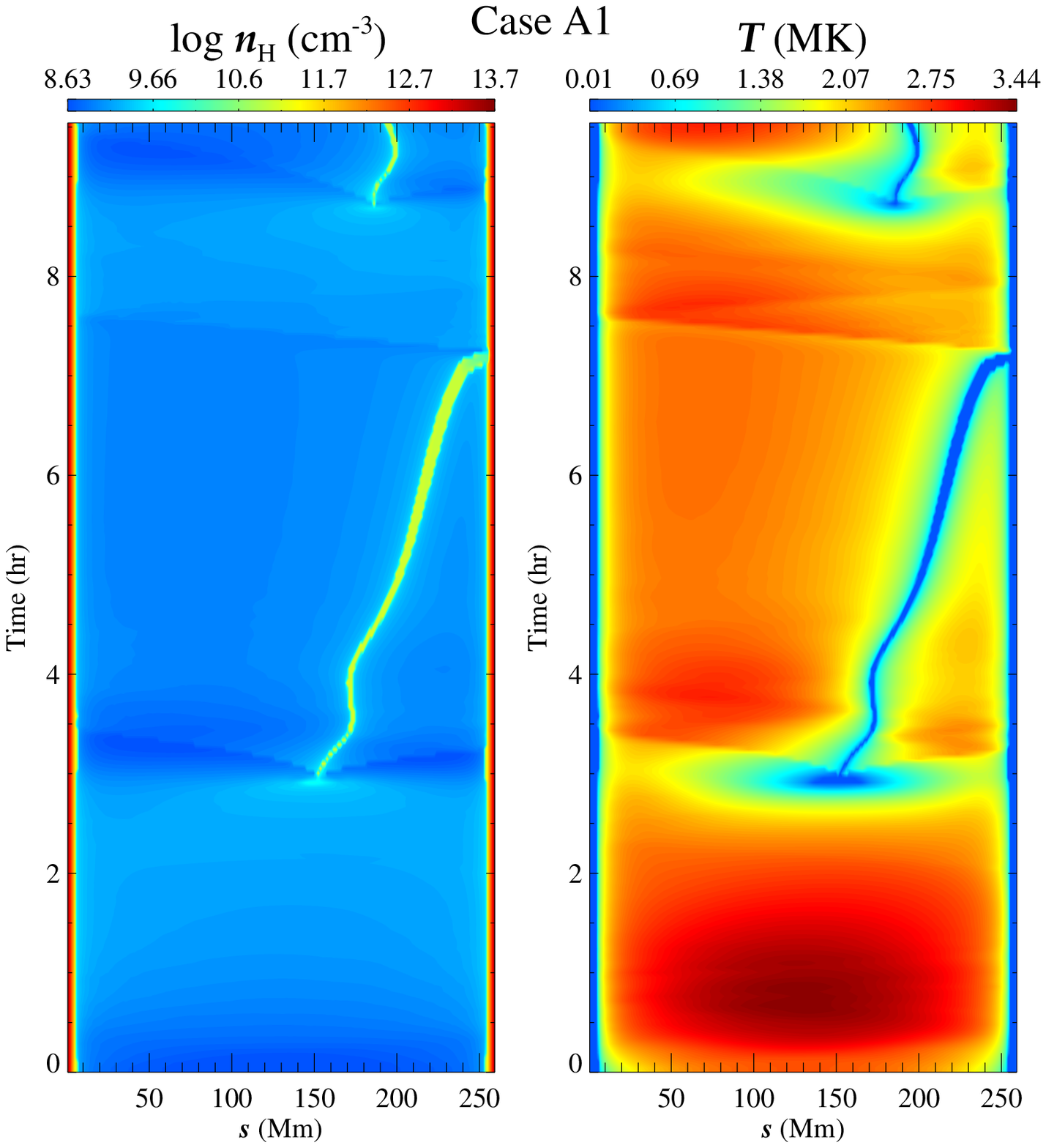}
\caption{Temporal evolution of the number density of hydrogen 
({\it left}) and the temperature ({\it right}) along the model loop 
in case A1. The loop footpoints are at $s=0$ and $260$ Mm, respectively,
and the center of the loop dip is at $s=130$ Mm.}
\label{fig:A1}
\end{figure}

\begin{figure}
\includegraphics[width=6.in]{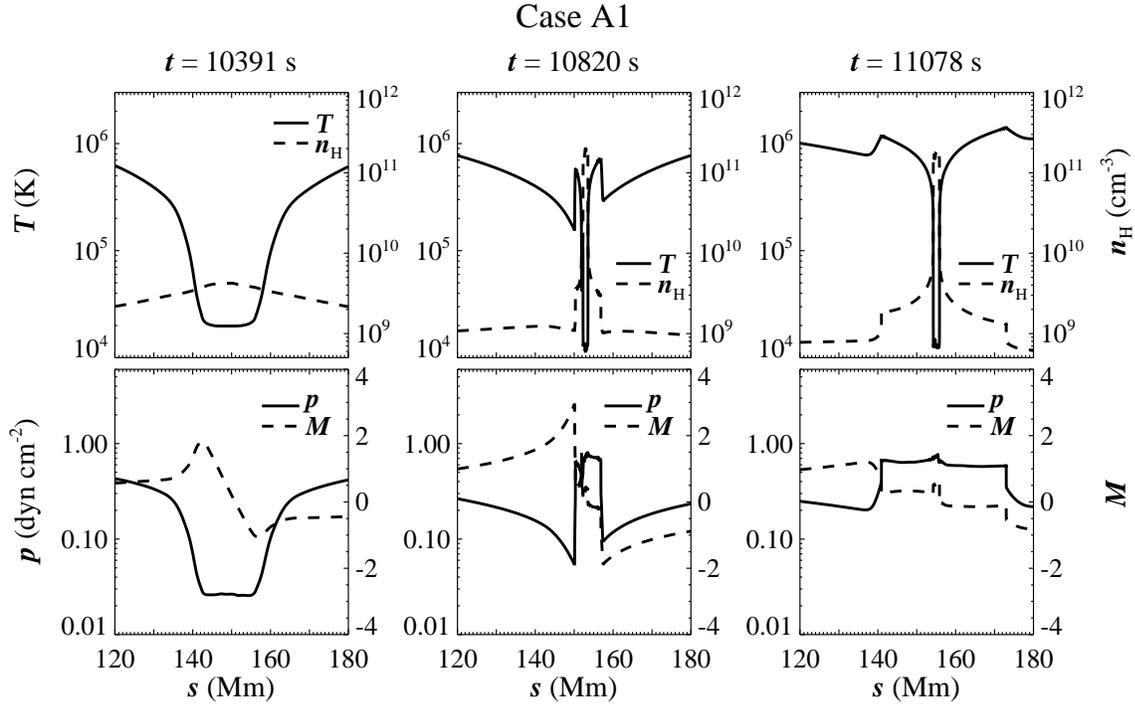}
\caption{Distributions of the temperature ({\it solid line, top}), the 
density ({\it dashed line, top}), the pressure ({\it solid line, bottom}), 
and the Mach number ({\it dashed line, bottom}) along the loop near the
midpoint, at three instants, namely $t=10391$ s ({\it left column}),
$t=10820$ s ({\it middle column}) and $t=11078$ s ({\it right column}),
in case A1.}
\label{fig:shocka}
\end{figure}

In case A1, we set $f=0.75$, and other parameters are the same as in
case S1. The details of the formation process are similar to case
S1, as illustrated by Figure \ref{fig:A1}, which depicts the time
evolution of the density ({\it left panel}) and temperature ({\it right
panel}) distributions. At $t=10690$ s, i.e., 2.97 hr, a condensation
with low temperature and high density is formed in the right part, i.e.,
the less heated part, of the magnetic dip, which is 22.5 Mm away from
the midpoint. The distributions of various quantities, e.g., the
temperature ($T$), the density ($n$), the pressure ($p$), and the in
situ Mach number ($M$), across the condensation segment at three times
are shown in Figure \ref{fig:shocka}. It is seen that as the convergent
inflows coalesce in the condensation segment, two rebound shock waves are
launched, propagating towards left and right footpoint, respectively. The
shock waves are reflected between each footpoint and the condensation
for several times before fading away, as also indicated by the 
sinusoidal pattern in the right panel of Figure \ref{fig:A1} near $t=
3.5$ hr. A significant difference from case S1, however, is  that upon
formation, the condensation has a velocity of $\sim 5$ km s$^{-1}$, moving
to the right, or the less heated side, as illustrated by Figure
\ref{fig:A1}. The rightward-moving condensation is accelerated to 15 km
s$^{-1}$ due to the pressure gradient on its two sides, but soon it is
dragged to decelerate by an inversion of the pressure gradient and even
falls back by 2 Mm. The pressure gradient inversion is caused by a
pressure increase on the right side of the condensation due to the
interaction of a shock wave after it is reflected at the right
footpoint, when the left-sided shock has not reached the left footpoint
yet. After the left-sided shock wave is reflected from the left
footpoint and catches the condensation, the pressure on the left side
of the condensation increases and exceeds the pressure on the right
side. The condensation is then pushed again by the pressure gradient to
move to the right with a velocity of $\sim 8$ km s$^{-1}$ until it
drains down to the right footpoint of the loop. During the travel, the
length of the condensation grows from 0.9 Mm to 7.9 Mm. When the
condensation impacts the chromosphere, the collision generates a
rebound shock wave, which is bounced back and forth between the two
footpoints of the loop as indicated by the sinusoidal pattern in the
right panel of Figure \ref{fig:A1} near $t=8$ hr. The total lifetime of
the condensation is $\sim 4.3$ hr. As the simulation goes on, the
formation and the drainage of condensation repeats, with a period of 5.5
hr, which is significantly shorter than the 22.8 hr simulated by
\citet{Karp06}. Since the radiative loss coefficent we used is generally 
$\sim 2$ times larger than theirs, the cooling is stronger and the 
condensations form faster, which shortens the period of the 
formation-drainage cycle.

\begin{figure}
\includegraphics[width=6.in]{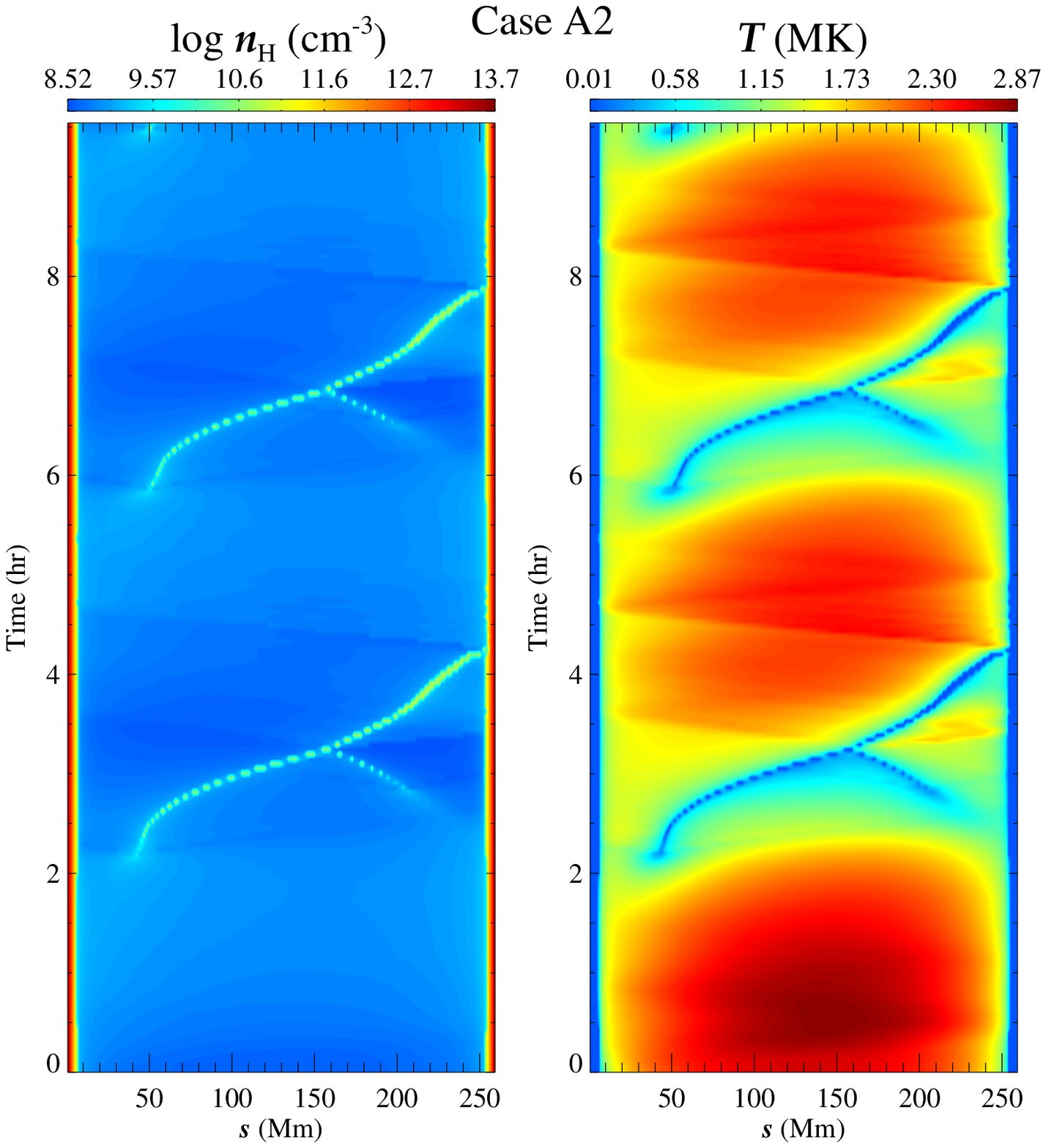}
\caption{Temporal evolution of the number density of hydrogen 
({\it left}) and the temperature ({\it right}) along the model loop 
in case A2. The two loop footpoints are at $s=0$ and $260$ Mm, 
respectively, and the center of the loop dip is at $s=130$ Mm.}
\label{fig:A2}
\end{figure}

As mentioned in subsection \ref{sym}, two condensations can be formed
when the heating scale length $\lambda$ is small. Therefore, in case
A2, we take $f=0.4$ and $\lambda=5$ Mm. As illustrated by Figure
\ref{fig:A2}, a condensation forms at $t=2.2$ hr near the left shoulder
of the magnetic dip ($s=42.1$ Mm) with an initial speed of 10 km
s$^{-1}$. It is accelerated during its travel towards the right part of
the loop, with its length growing up to 5.8 Mm. At $t=2.84$ hr (38 min
later), a second condensation, which is smaller than the first one, is
formed near the right shoulder of the magnetic dip ($s=205.5$ Mm) with
an initial speed of 24 km s$^{-1}$ towards the left. The left and the
right condensation segments are accelerated for 24 min to 60 km s$^{-1}$
and 50 km s$^{-1}$, respectively, and then collide at $s=158$ Mm (at the
right part of the loop). Two shock waves are generated by the collision,
which are then reflected back and forth between the condensation and
each footpoint, as revealed by the sinusoidal pattern in the right panel
of Figure \ref{fig:A2} near $t=3.5$ hr. After the collision, the two
condensation segments merge into one, with a length of 2.3 Mm. The
coalesced condensation moves to the right with an initial velocity of 32
km s$^{-1}$. It is decelerated to 16 km s$^{-1}$ at $s=216$ Mm, and then
accelerated to 24 km s$^{-1}$ with a length of 6 Mm before it drains
down to the right footpoint of the loop. The deceleration and
acceleration of the condensation is mainly due to the inversion of the
pressure gradient caused by the same reason as discussed in case A1, 
while the gravity effect is small in this shallow dip configuration. 
The falling down of the condensation excites a shock wave, which is 
trapped to propagate back and forth in the whole loop as indicated by 
the right panel of Figure \ref{fig:A2} near $t=5$ hr. As time goes on, 
such a formation, coalescence, and drainage of condensations repeat 
with a period of 3.6 hr, which is shorter than that in case A1. 

As an extreme case, we perform a simulation with $f=0$, i.e., the 
localized heating is introduced at the left footpoint only. It is found
 that no condensation forms in the loop. Instead, we get steady flows 
along the coronal loop, consistent with previous works
\citep[e.g.,][]{Pats04}.

\section{Discussions}\label{discuss}

\subsection{Thermal Instability}

\citet{Park53} proposed that some solar activities, such as filaments,
can be formed by thermal instability. He derived a criterion for 
thermal instability on the basis of an analysis of the energy equation 
alone. \citet{Fiel65} made a detailed research of the thermal 
instability for an infinite, uniform, static plasma in initial thermal
equilibrium. He pointed out that the criterion given by Parker is based
on the isochoric assumption, i.e., the density is constant in the whole
region, which is not compatible with the force equation since the
cooling would lead to pressure deficit, which would destroy the initial
force balance. He derived an isobaric criterion for the thermal
instability, which is consistent with the force equation. The thermal 
instability was further studied by many other colleagues 
\citep[e.g.,][and references therein]{vand91,Meer96}. It was pointed out
that these modes are the marginal entropy modes which are advected with
the local flow velocity \citep{Goed10}, driven unstable by non-adiabatic
processes. The different criteria may be applicable for different
astrophysical environments.

According to our simulations, as mentioned in \S\ref{sym}, during the 
catastrophic cooling stage, the temperature and the pressure drop 
rapidly, while the density increases only a little at this stage.
Significant density enhancement occurs $\sim 3$ min after the
catastrophic cooling. Therefore, in the simulated coronal loop,
the thermal catastrophe is more isochoric than isobaric. So we use
the isochoric thermal instability criterion derived by \citet{Park53} as
follows:

\begin{equation}\label{crit}
C\equiv k^2- \frac{1}{\kappa} \left(\frac{\partial H(s)}{\partial T}
-\frac{\partial R}{\partial T}\right) < 0
\end{equation}
where $k$ is the wave number of the perturbations, $\kappa$ is the heat 
conduction coefficient, and $R=n_{\rm H} n_{\rm e} \Lambda(T)$ is the
radiative loss. The heat conduction introduces a stabilizing effect.
We
numerically calculate $\partial R /\partial T$, using the central
difference scheme. $\partial H(s) /\partial T$ is zero since the 
heating depends only on space in our simulations. Perturbations with 
any resolvable wavelength exist in the simulations. According to Figure
\ref{fig:conden}, the cool region has a width of $\sim 10$ Mm, 
therefore, we take the wavelength of the temperature perturbation as 20 
Mm in order to quantify $k$. For small $k$, the occurrence of the 
thermal instability is mainly determined by the sign of 
$\partial R /\partial T$.

\begin{figure}
\includegraphics[width=6.in]{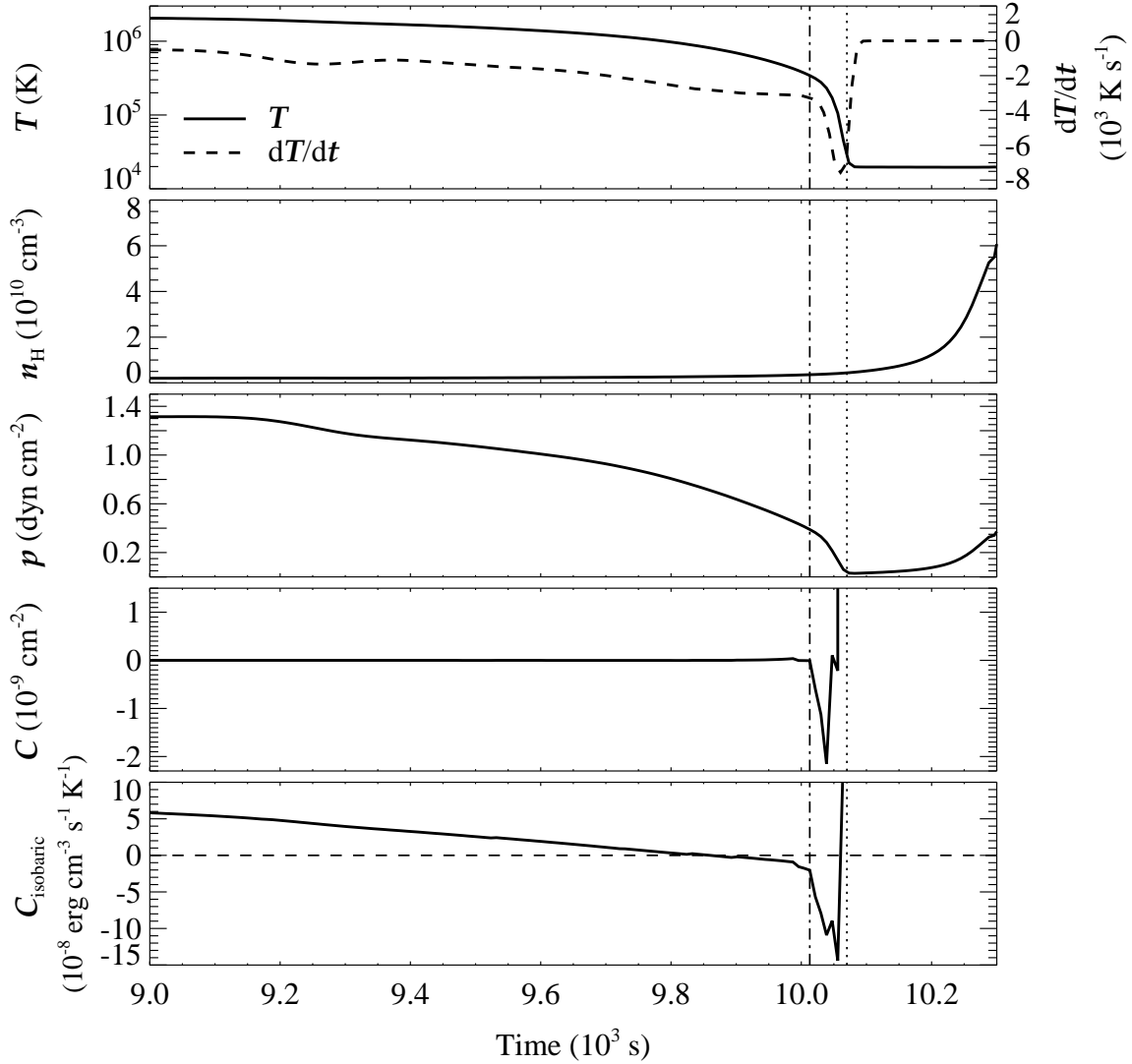}
\caption{Temporal evolution of the temperature, the density, the 
pressure, the isochoric criterion $C$, and the isobaric criterion 
$C_{\rm isobaric}$ at the midpoint ({\it solid lines}), as well as 
the time derivative of the temperature ({\it dashed line, top panel}). 
Note that the vertical dotted dashed line denotes $t=10013$ s when the 
thermal instability begins, and the vertical dotted line denotes the 
end of catastrophic cooling, $t=10070$ s.}
\label{fig:crit}
\end{figure}

Taking case S1 as an example, we plot the temporal evolution of the
temperature, the density, the pressure, and the isochoric criterion $C$
at the loop midpoint in Figure \ref{fig:crit}. Since the initial
temperature is 2.63 MK, which corresponds to a negative $\partial
R /\partial T$, $C=-1.4\times 10^{-15}$ cm$^{-2}$ is negative, but very
close to 0. Besides, the cooling timescale at this stage is $\sim 10^4$
s. Therefore, the early evolution is dominated by the localized heating
and chromospheric evaporation. As more mass is filled into the corona,
radiation is enhanced gradually, which becomes overwhelming over the
heating after $t=2664$ s. The temperature keeps decreasing slowly then.
From $t=9850$ s to $t=9994$ s, $C$ becomes positive for a short interval
since $T$ falls in the range where $\partial R /\partial T$ is
positive. After $t=10013$ s, $C$ drops down drastically to $-1.2\times
10^{-9}$ cm$^{-2}$. Simultaneously, the temperature $T$, along with the
gas pressure, begins to decrease catastrophically, as indicated by the
time derivative of $T$, i.e., the dashed line in the top panel of Figure
\ref{fig:crit}. That is to say, the thermal instability occurs.  The
temperature drops from $3.4 \times 10^5$ K to 20000 K in 60 s. However,
the density increases by only 20\% during this time. Note that $C$ 
becomes positive out of the plotting range after the catastrophic
cooling, corresponding to a thermally stable state. Three minutes later,
i.e., at $t=10282$ s, the density increases drastically, and a
condensation is then formed. We conclude that the isochoric thermal
instability may explain the catastrophic cooling. 
Such a delay is probably due to the difference between the kinematic
timescale and the radiative timescale. It takes an extra 3 min for the
plasma to accumulate in the cooling region under the pressure gradient 
driving.

It might be interesting to check the criterion of the isobaric thermal
instability. According to Equation (25) of \citet[][see also 
\citealt{vand91}]{Fiel65}, the criterion for the thermal instability 
in the isobaric case is expressed as:

\begin{equation}
C_{\rm isobaric}\equiv \rho \left(\frac{\partial \mathcal{L}}{\partial
        T}\right)_{\rho} - \frac{\rho ^2}{T}\left(\frac{\partial
        \mathcal{L}}{\partial \rho}\right)_T+k^2 \kappa < 0
\end{equation}
\noindent
where $\mathcal{L}=(n_{\rm H} n_{\rm e} \Lambda(T)-H(s))/\rho$ is the
generalized heat-loss function. For the perturbations with a wavelength
of 20 Mm, we calculate the $C_{\rm isobaric}$ at the midpoint of the
loop and plot its temporal evolution in the bottom panel of Figure
\ref{fig:crit}. The time when it turns from positive to negative is 
well before the onset time of the catastrophic cooling. We try 
many other perturbation wavelengths, and it is found that the isobaric 
criterion is crucially dependent on the perturbation wavelength while the
isochoric criterion is not. Therefore, we conclude that the isobaric 
thermal instability is not appropriate to explain the catastrophic 
cooling during the condensation formation in the solar corona.

\subsection{Is Continued Heating Necessary?}\label{nece}

\begin{figure}
\includegraphics[width=6.in]{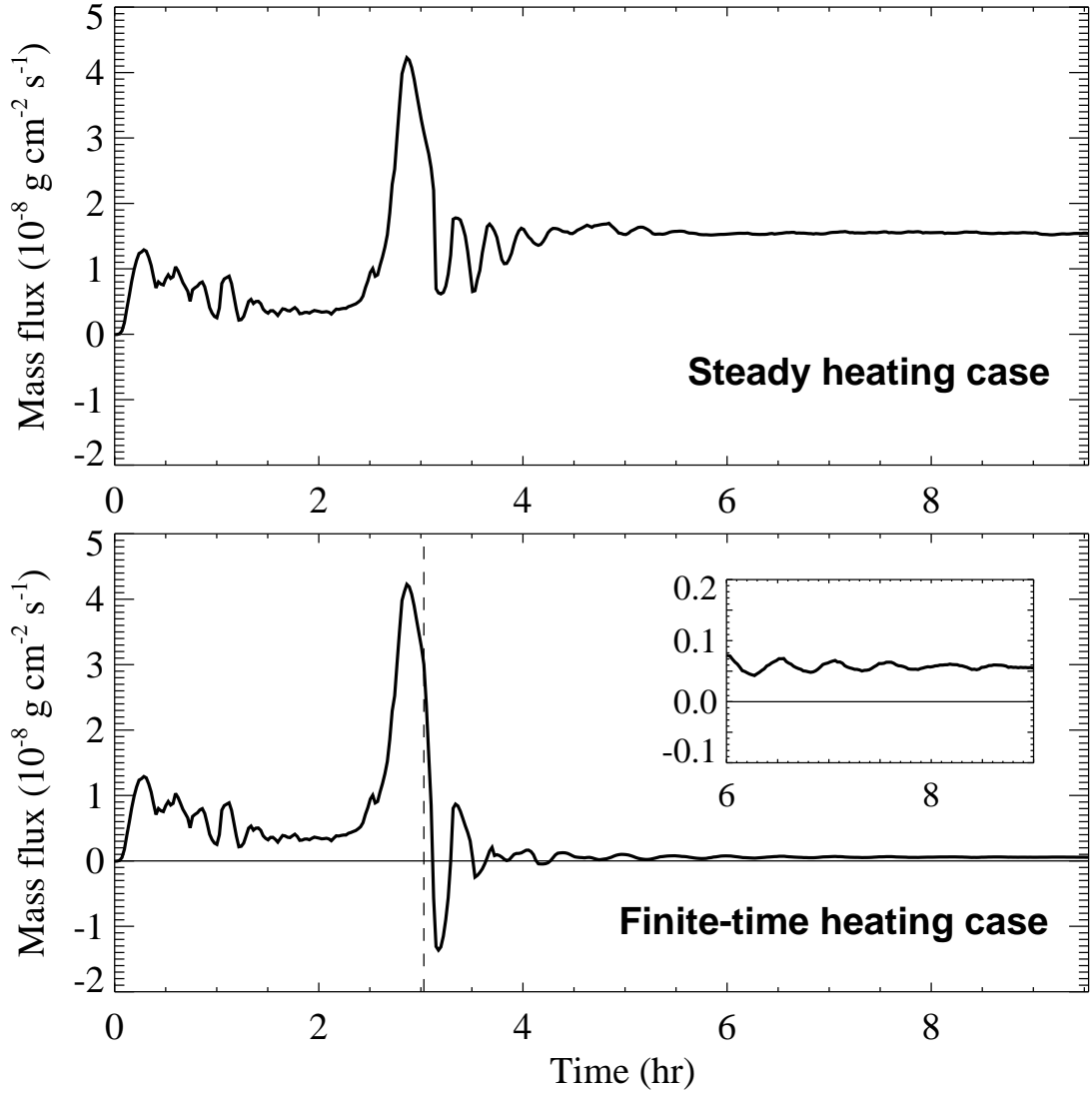}
\caption{Evolutions of the mass flux from two shoulders of the magnetic
dip to the midpoint of the loop in the steady heating case ({\it top 
panel}) and the finite-time heating case ({\it bottom panel}). Note 
that the dashed line at $t=3.03$ hr in the bottom panel marks the 
moment when the localized heating is switched off, and the inset shows 
the zoom-in view of the curve from $t=6$ hr to 9 hr. The horizontal 
thin line indicates the zero level.}
\label{fig:mflux}
\end{figure}

As mentioned in \S\ref{intro}, it has been demonstrated that the extra
heating localized in the low atmosphere would drive chromospheric
evaporation flows, leading to the plasma condensation in the corona due
to thermal instability or loss of thermal equilibrium. In the previous
studies, the localized heating is either continuous \citep{Anti99} or
intermittent \citep{Karp08}. In the steady heating case, the 
condensation can form and grow rapidly, as also demonstrated in this 
paper. In the successive impulsive heating case, it was found that 
a condensation can also form steadily when the average interval between 
heating pulses is less than the coronal radiative cooling time 
\citep[$\sim 2000$ s,][]{Karp08}. From the theoretical point of view,
the localized strong heating may be due to low atmospheric activities,
such as chromospheric reconnection. It is quite possible that such a
heating event has a finite lifetime and might not show up again at the 
footpoint of one flux tube. Therefore, it is interesting to see the 
response of the coronal loop to a single heating event. To do that, we
 make a numerical experiment, and stop the localized heating 8 minutes 
after the condensation is formed at $t$=2.87 hr in the symmetric case 
S1, which means there is only background heating $H_0(s)$ in the heating
term $H(s)$ in Equation \ref{eq3}, which changes slightly with the 
distance along the loop. Note that, in order to make the evolution more 
smooth, the heating is turned off linearly over 1000 s. We find that 
after the heating ceases, there is still mass upflows from the footpoint 
to the coronal portion of the loop. For comparison, Figure 
\ref{fig:mflux} plots the evolutions of the mass flux from the two
shoulders of the magnetic dip to the condensation segment in the steady
case ({\it top panel}) and the finite-time heating case ({\it bottom 
panel}). It is seen that after the condensation is formed at $t=2.87$ 
hr, the mass flux oscillates heavily and then maintains at a level of 
$1.5\times 10^{-8}$ g cm$^{-2}$ s$^{-1}$ in the steady heating case. 
However, in the finite-time heating case, the mass flux drops abruptly,
but then still maintains at a level of $5.6\times 10^{-10}$ g cm$^{-2}$
 s$^{-1}$ even after the localized heating is removed permanently. That
is to say, the mass flux decreases by 27 times, but does not vanish.
Such a mass flux corresponds to a growth rate of the condensation length
of 230 km hr$^{-1}$, which decreases by 6.6 times compared to the steady
heating case. The reason why the growth rate does not decrease
proportionally with the mass flux is that the plasma density of the
condensation is reduced after the heating is switched off.

Therefore, it seems that the evaporation-driven condensation can serve
as the trigger of the formation of the filament threads, and there
exists a condensation instability. Once the condensation is formed in a
small segment near the dip of a coronal loop by the evaporation flow,
the condensation will grow, and there is continual mass supply siphoned
from the chromosphere, although the growth rate is $\sim 6.6$ times
smaller than in the steady heating case.

\begin{figure}
\includegraphics[width=6.in]{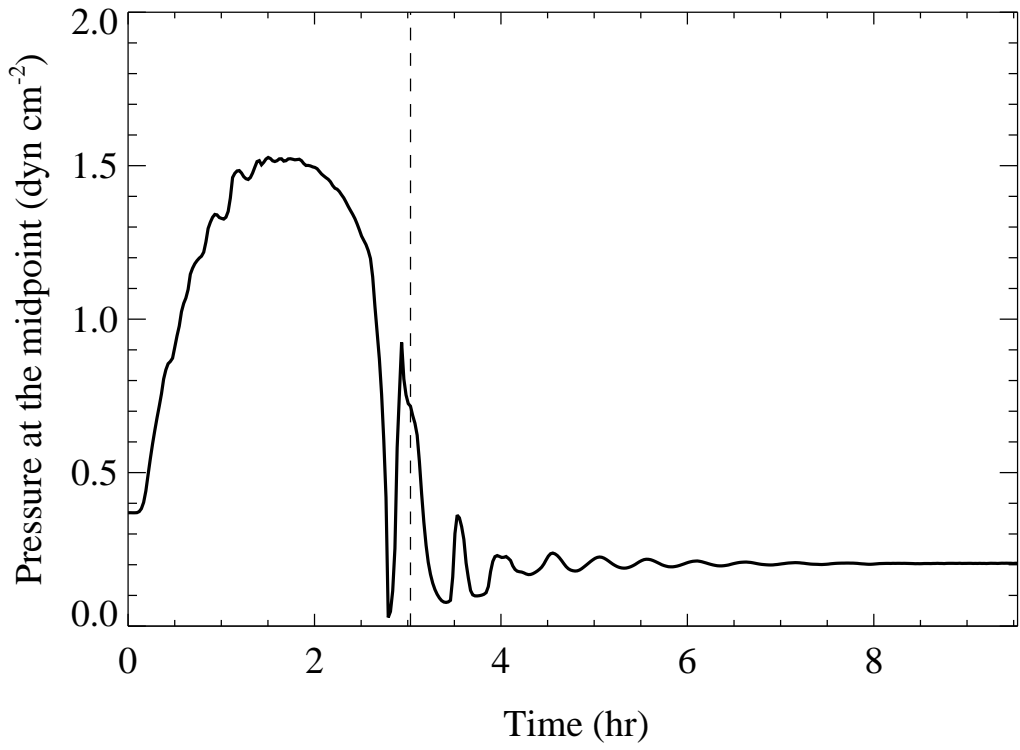}
\caption{Evolution of the gas pressure at the midpoint of the loop
in the finite-time heating case. The dashed line indicates $t=3.03$ 
hr when the localized heating is switched off.}
\label{fig:press}
\end{figure}

In order to understand the mechanism of the spontaneous siphon flow
after the localized heating is halted, we plot in Figure 
\ref{fig:press} the time evolution of the gas pressure at the midpoint
 of the loop, where the condensation is located. It is revealed that 
after the localized heating is turned off gradually, the pressure at 
the loop midpoint drops down from $t=3.03$ hr. After several hours of 
small-amplitude oscillations, the pressure remains at 0.2 dyn cm$^{-2}$,
which is about half of its initial value 0.37 dyn cm$^{-2}$ at $t=0$.
 Therefore, our simulation result indicates that after thermal
instability and plasma condensation, 
the gas pressure of the cold plasma is reduced, compared to the hot
 plasma at the same site in the initial hydrostatic state, which leads
 to a pressure gradient along the loop. It is such a pressure gradient
 that drives the spontaneous siphon flow, which makes the condensation
continue to grow even after the localized heating is switched off.

\subsection{Stability to {\it p}-mode wave perturbations}\label{wave}

\begin{figure}
\includegraphics[width=6.in]{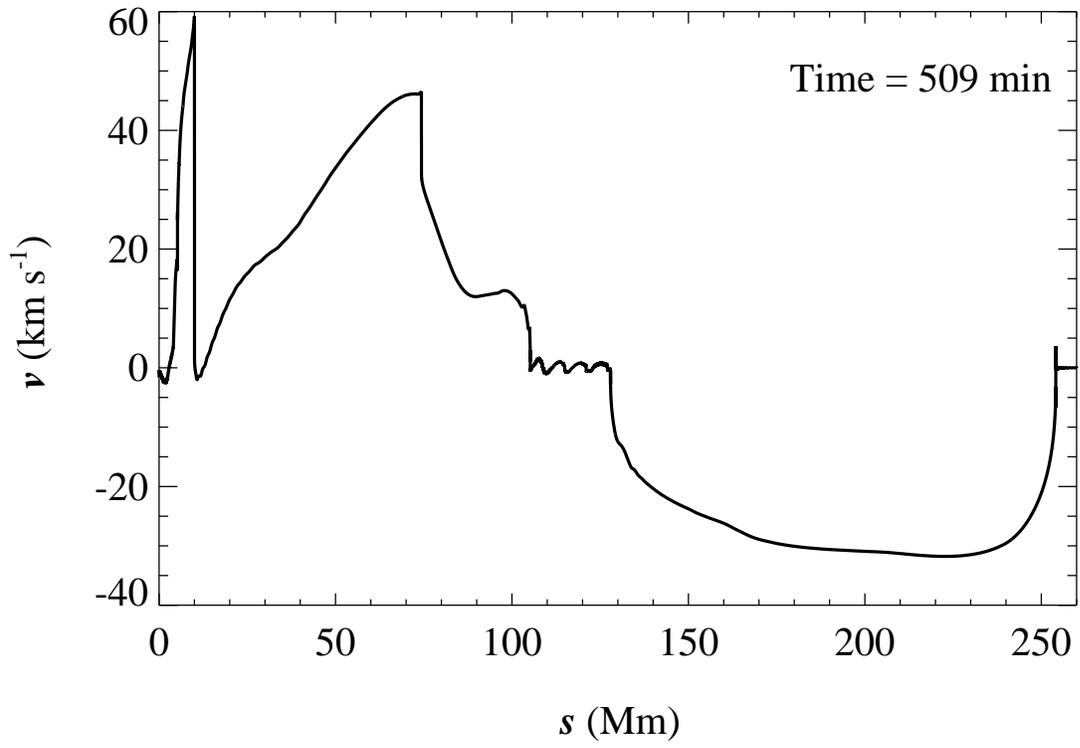}
\caption{Velocity distribution along the loop at $t=509$ min in case
D1.}
\label{fig:vt}
\end{figure}

\begin{figure}
\includegraphics[width=6.in]{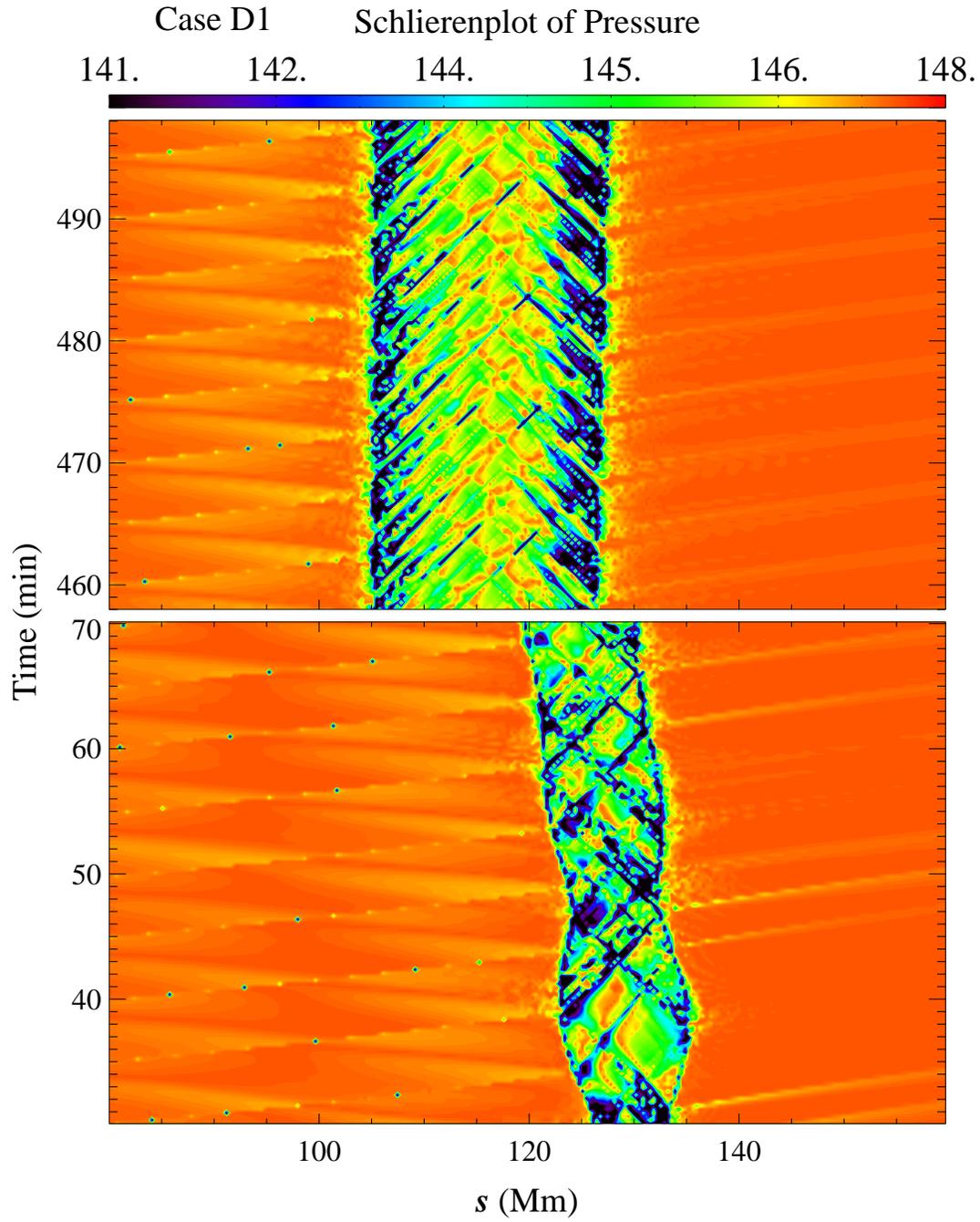}
\caption{Temporal evolution of Schlierplot of the pressure in the
region around the condensation. The lower part shows the initial
drifting phase and the upper part shows the final stable phase.}
\label{fig:spt}
\end{figure}

While the previous section demonstrated that prominence growth will 
continue even after finite-time localized heating, another aspect 
worth studying is the fate of the prominence condensations subjected to 
wave buffeting. Since 5-minute solar {\it p}-mode oscillations are 
ubiquitous in the photosphere, it is relevant to investigate whether 
prominences will be influenced by wave driving, and how they channel 
(linear) wave modes. Since there is large interest in prominence 
seismology \citep[see, e.g., the review by][]{Mack10}, we now 
investigate how filaments, once formed, behave under {\it p}-mode 
wave driving, which is denoted as case D1. Case D1 is based on the
simulation results of the symmetric case S1, with an initial state
taken from 7.1 hours in the evolution shown in Figure~\ref{fig:S1}. We
then introduce a sinusoidal velocity perturbation with the amplitude of
1 km s$^{-1}$ and the period of 5 minutes. We add this perturbation only
at the left footpoint of the loop. The {\it p}-mode waves propagate
upward through the transition region into the corona and steepen into
shocks. A snapshot of the velocity distribution after 509 minutes is 
shown in Figure \ref{fig:vt}, where two shock fronts can be identified
to the left of the central condensation. The shocks damp in the 
corona while propagating with a speed of 213 km s$^{-1}$, which is 
close to the local sound wave speed. These shocks hit the 
condensation and penetrate into it with little reflection, becoming 
ordinary linear sound waves while propagating through the filament. 
When these sound waves hit the other boundary of the condensation, 
they are mainly reflected, with a weak leakage out to the right part of 
the loop. This is best visualized using a Schlieren plot 
of the pressure, as shown in Figure \ref{fig:spt}. This Schlieren plot 
of the pressure zooms in around the condensation, and 
actually quantifies the local value of 
$\exp(-0.01[\left|\nabla p \right|-500])$. The bottom part of 
Figure ~\ref{fig:spt} shows the first series of shocks hitting the 
condensation. One notes that the overall thermodynamical changes 
induce a leftward drift of the central prominence. The symmetry is 
hereby broken due to our asymmetric driving. During the leftward drifting,
the filament thread becomes longer due to the chromospheric evaporation
as in case S1. The top panel of Figure~\ref{fig:spt} shows that ultimately
the filament settles down at a quasi-permanent location determined by the 
overall pressure balance, as altered by the periodic driving. The wave 
mode reflections and transmissions at the left and right edges of the 
filament thread can be clearly detected. The different slope of 
the wave fronts within the filament is due to the lower sound 
speed there. All these waves, either inside or outside the filament thread,
have a 5-min period.  The impinging waves are 
clearly nonlinear, while the internal wave modes are primarily linear, 
and the transmitted waves are further attenuated. During the entire 
period simulated, the condensation maintains thermally stable under the 
perturbation and energy damping from {\it p}-mode waves.

\subsection{Summary of new findings}\label{new}

Our model is appropriate for low-lying, shallowly-dipped loops, which 
have not been studied before. Such shallow dip configurations 
facilitate higher speeds for displacing the condensations. As
advocated by studies of hydrostatic coronal loops by \citet{Asch02},
our background heating uses an exponential height dependence, differing
from the uniform prescription used in previous works. In case S1, the 
great details of the condensation process are shown for the first time.
Our parameter survey finds new complicated cycles of paired filament 
formation, with high-speed converging condensations in case A2, with 
strong-asymmetry and short-scale localized heating. With the improved 
cooling table, we allow for stronger radiative cooling and therefore 
find shorter formation time scales compared to previous works. 
Besides continuous 
localized heating, the case with finite duration, which is more realistic
in the solar atmosphere, is investigated for the first time, which
indicates that the extra strong heating is unnecessary to maintain growth 
of condensations after their onset. We studied the dependence of the 
formation process for a large parameter range of the heating scale length
$\lambda$ and the heating amplitude $E_1$. It is found that shorter 
$\lambda$ or stronger $E_1$ can make condensations form earlier, say, 
$\sim$2 hr after the introduction of the localized heating. Shorter 
 $\lambda$ also leads to faster growth of the condensation. 
Moreover, the effect of $p$-mode waves is studied for the first time in 
this context.

\section{Conclusions}\label{con}

It has been suggested that localized heating in the chromosphere can
drive plasma evaporation into the corona, and form plasma condensation
through thermal instability or loss of thermal equilibrium. In order to
investigate the details of this process, in this paper we performed 1D
radiative hydrodynamic simulations in a magnetic loop, where heat
conduction, radiative losses, and heating terms are included in the
energy equation. The main results can be summarized as follows:

(1) The cold condensation formation can be divided into three stages, 
namely, a thermal rearrangement stage, a thermally unstable stage, and a
kinematic stage. In the first stage, as more chromospheric mass is
evaporated into the corona, the radiative cooling is enhanced, so the
temperature decreases slightly and steadily. In the second stage, the
criterion of the isochoric thermal instability is satisfied, and both
the plasma temperature and pressure drop down rapidly. They reach their
minimum in $\simeq 1$ min. In the third stage, strong inflows due to huge
pressure gradient are driven towards the cold region. They collide,
launching shock waves after forming a condensation in $\sim 3$ min. The
3-min delay of the condensation formation with respect to the 
catastrophic cooling relates to the longer kinematic timescale of the 
plasma than the cooling timescale.

(2) When the localized steady heating at the two footpoints is
symmetric, one cold plasma condensation forms at the midpoint of the
loop, and grows steadily. Our parameter survey indicates that the onset
time of the condensation varies from $t=$2 hr to $t=$5 hr, and the mean
growth rate varies from 800 km hr$^{-1}$ to 4000 km hr$^{-1}$, depending
on the amplitude and the scale length of the heating function.

(3) When the localized heating at the two footpoints is weakly
asymmetric, also one condensation forms. However, it is shifted from the
midpoint of the loop towards the less heated footpoint. It moves with a
velocity of $\sim 15$ km s$^{-1}$, and then drains down to the less 
heated footpoint. When the heating at the two footpoints is strongly
asymmetric and the heating scale length is short, two condensation 
segments form at the two shoulders of the magnetic dip, successively.
The two segments move towards each other with a relative
velocity of up to 50 km s$^{-1}$, which might account for the
counter-streaming found in observations. The two segments finally merge
into one segment, which moves and then drains down to the less heated
footpoint with a velocity up to 24 km s$^{-1}$.

(4) As an extreme case, when heating is localized at one footpoint, no
plasma condensation can be formed, and only steady flow is obtained
along the coronal loop, as also demonstrated by \citet{Pats04}.

(5) It is found that once formed the condensation can grow even if the
localized heating ceases, though the growth rate of the condensation
length, $\sim 230$ km hr$^{-1}$, is $\sim 6.6$ times smaller than in the
steady heating case. Our research suggests that there exists a
condensation instability, i.e., after thermal instability and plasma
condensation, the gas pressure becomes reduced, and the pressure
gradient drives spontaneous siphon flows from the chromosphere to the
corona, which helps the further growth of the condensation.

(6) The plasma condensation maintain its stability and keeps
growing, even when {\it p}-mode waves propagate through it. The fact
that waves can be transmitted through the filaments is relevant for 
prominence seismology, although our results are restricted to
longitudinal acoustic waves.

It should be noted that our assumptions, such as the fully ionized
plasma and the optically thin radiative cooling, may not be appropriate
to investigate details of the dense partially ionized plasmas in
filaments. The ionization and radiation transfer in the optically thick
plasma should be considered in the future to reproduce the observational
characteristics of filaments. The effects of the {\it p}-mode waves from
the photosphere on the formation of filaments will also be investigated.
Multi-dimensional MHD simulations are also planned in order to
understand the coupling between the plasma and the magnetic field.

\acknowledgments
The authors thank Z. Meliani, T. Berger, Y. Guo, and M.D. Ding for 
discussions and thank J. Colgan for offering his radiative cooling data.
The visit of C. Xia to K.U. Leuven is supported by the Graduate School 
of Nanjing University and Center for Plasma Astrophysics in K.U. Leuven. 
The research is supported by the Chinese foundations NSFC (11025314, 
10403003, 10933003, and 10673004) and 2011CB811402. These results are 
supported by the project GOA/2009/009 (K.U.Leuven). For some simulations 
we used the infrastructure of the VSC - Flemish Supercomputer Center, 
funded by the Hercules foundation and the Flemish Government, department 
EWI.

\clearpage

\clearpage

\clearpage

\clearpage

\clearpage

\clearpage

\clearpage

\clearpage

\clearpage

\clearpage

\clearpage

\clearpage

\clearpage

\clearpage

\clearpage

\clearpage

\clearpage

\clearpage

\clearpage

\begin{table}
\begin{center}
\caption{The parameters and results of the typical cases\label{tab1}}
\begin{tabular}{ccccccc}
\\ \tableline\tableline
case& $\lambda$ & $f$ & $E_1$ & Mean Growth Rate &Onset Time&Segment\\
    &   (Mm)    &   &(erg cm$^{-3}$
                     s$^{-1}$) & (km hr$^{-1}$)&  (s)     &  Number\\
\tableline
S1  &    10     & 1  &   0.01    &   1928      &   10340  & 1\\
S2  &    5      & 1  &   0.01    &   3228      &   7729   & 1\\
S3  &    10     & 1  &   0.02    &   1505      &   9618   & 1\\
A1  &    10     &0.75&   0.01    &   1842      &   10648  & 1\\
A2  &    5      &0.4 &   0.01    &   3079      &   7900   & 2\\
\tableline
\end{tabular}
\end{center}
\end{table}

\begin{table}
\caption{A list of parameters in simulations of radiative condensation
\label{tab2}. These relate to the overall assumed loop geometry,
quantifying the loop length, the presence of a dip, the length of the
loop legs, and the height of the chromosphere.}
\begin{tabular}{lccccc}
\\ \tableline\tableline
Reference& $L$ & $D$ & Vertical Leg & $s_{tr}$&Cross Section \\
  &Mm&Mm & Mm & Mm  & (Non)Uniform  \\
\tableline
\citet{Anti99} & 220 & 5   & 10 &10   & U \\
\citet{Anti00} & 320 & 5   & 60 &50   & U \\
\citet{Karp01} & 340 & no  & 60 &60   & U \\
\citet{Karp03} & 420 &15,10& 75 &60   & U \\
\citet{Mull03} & 10  & no  & 1  &1.6  & U \\
\citet{Mull04} & 100 & no  & 1  &1.6  & U \\
\citet{Karp05} & 405 & 20  & 60 &60   & N \\
\citet{Karp06} & 405 & 20  & 75 &60   & U \\
Karpen et al. (2008)&405&20& 75 &60   & U \\
\citet{Klim10} & 205 & no  & 50 &50   & U \\
Our Cases      & 260 & 0.5 & 5  &6    & U \\
\tableline
\end{tabular}
\end{table}

\begin{table}[htbp]
\centering
\begin{threeparttable}
\caption{A list of parameters in simulations of radiative
condensation\label{tab3}. These relate to the heating adopted,
and to the radiative cooling prescriptions.}
\begin{tabular}{lcccccc}
\\ \tableline\tableline
Reference& $E_0$ & $E_1$ & $f$ &$\lambda$&Type&Radiation \\
  &erg cm$^{-3}$ s$^{-1}$& erg cm$^{-3}$ s$^{-1}$ & &Mm&S/I/F\tnote{1}  & \\
\tableline
\citet{Anti99} & 1.5e-5& 1.e-3 & 1  &  10 & S & Old\tnote{2} \\
\citet{Anti00} & 1.5e-5& 1.e-3 &0.75&  10 & S & Old \\
\citet{Karp01} & 1.5e-4& 1.e-3 &0.75&  10 & S & Old \\
\citet{Karp03} & 1.5e-4& 1.e-2 &0.75&  10 & S & Old \\
\citet{Mull03} & no    & 1.2e-3& 1 & 1.25 & S & IE\tnote{3} \\
\citet{Mull04} & no    & 1.2e-3& 1 & 5,3,2& S & IE \\
\citet{Karp05} & 1.5e-4& 1.e-2 & 0.75& 10 & S & KR\tnote{4} \\
\citet{Karp06} & 1.5e-4& 2.e-2,1.e-2&0.75&5,10&S&KR\\
Karpen et al. (2008) & 1.5e-4& 1.e-2 & 0.75& 5,1 &I&KR\\
\citet{Klim10} & 6.e-4 & 8.e-2 & 0.5,0.75,0.9&5&S&KR\\
Symmetric Cases&3.e-4 & 5.e-3$\sim$0.2&1&3$\sim$20&S/F&Colgan\\
Asymmetric Cases & 3.e-4 & 1.e-2 & 0.4,0.75& 5,10 & S & Colgan \\
\tableline
\end{tabular}
\begin{tablenotes}
 \item[1] Steady/Impulsive/Finite heating
 \item[2] A simple piecewise radiative loss function which is an order 
          of magnitude smaller than the updated Klimchuk-Raymond (KR) version
 \item[3] Radiative loss included by solving Ionization Equations
 \item[4] Klimchuk-Raymond radiative loss function
\end{tablenotes}
\end{threeparttable}
\end{table}


\begin{thebibliography}{55}
\expandafter\ifx\csname natexlab\endcsname\relax\def\natexlab#1{#1}\fi

\bibitem[{Antiochos et~al.(2000)Antiochos, MacNeice, \& Spicer}]{Anti00}
{Antiochos}, S.~K., {MacNeice}, P.~J., \& {Spicer}, D.~S. 2000, \apj, 536, 494

\bibitem[{{Antiochos} {et~al.}(1999){Antiochos}, {MacNeice}, {Spicer}, \&
  {Klimchuk}}]{Anti99}
{Antiochos}, S.~K., {MacNeice}, P.~J., {Spicer}, D.~S., \& {Klimchuk}, J.~A.
  1999, \apj, 512, 985

\bibitem[{{Aschwanden}(2001)}]{Asch01}
{Aschwanden}, M.~J. 2001, \apj, 560, 1035

\bibitem[{{Aschwanden} \& {Schrijver}(2002)}]{Asch02}
{Aschwanden}, M.~J. \& {Schrijver}, C.~J. 2002, \apjs, 142, 269

\bibitem[{{Aulanier} {et~al.}(1998){Aulanier}, {Demoulin}, {van
  Driel-Gesztelyi}, {Mein}, \& {Deforest}}]{Aula98}
{Aulanier}, G., {Demoulin}, P., {van Driel-Gesztelyi}, L., {Mein}, P., \&
  {Deforest}, C. 1998, \aap, 335, 309

\bibitem[{{Berger} {et~al.}(2008){Berger}, {Shine}, {Slater}, {Tarbell},
  {Title}, {Okamoto}, {Ichimoto}, {Katsukawa}, {Suematsu}, {Tsuneta}, {Lites},
  \& {Shimizu}}]{Berg08}
{Berger}, T.~E., {Shine}, R.~A., {Slater}, G.~L., {Tarbell}, T.~D., {Title},
  A.~M., {Okamoto}, T.~J., {Ichimoto}, K., {Katsukawa}, Y., {Suematsu}, Y.,
  {Tsuneta}, S., {Lites}, B.~W., \& {Shimizu}, T. 2008, \apjl, 676, L89

\bibitem[{{Chae} {et~al.}(2001){Chae}, {Wang}, {Qiu}, {Goode}, {Strous}, \&
  {Yun}}]{Chae01}
{Chae}, J., {Wang}, H., {Qiu}, J., {Goode}, P.~R., {Strous}, L., \& {Yun},
  H.~S. 2001, \apj, 560, 476

\bibitem[{{Choe} \& {Lee}(1992)}]{Choe92}
{Choe}, G.~S. \& {Lee}, L.~C. 1992, \solphys, 138, 291

\bibitem[{{Colgan} {et~al.}(2008){Colgan}, {Abdallah}, {Sherrill}, {Foster},
  {Fontes}, \& {Feldman}}]{Colg08}
{Colgan}, J., {Abdallah}, Jr., J., {Sherrill}, M.~E., {Foster}, M., {Fontes},
  C.~J., \& {Feldman}, U. 2008, \apj, 689, 585

\bibitem[{{Dahlburg} {et~al.}(1998){Dahlburg}, {Antiochos}, \&
  {Klimchuk}}]{Dahl98}
{Dahlburg}, R.~B., {Antiochos}, S.~K., \& {Klimchuk}, J.~A. 1998, \apj, 495,
  485

\bibitem[{{Engvold}(2004)}]{Engv04}
{Engvold}, O. 2004, in IAU Symposium, Vol. 223, Multi-Wavelength Investigations
  of Solar Activity, ed. {A.~V.~Stepanov, E.~E.~Benevolenskaya, \&
  A.~G.~Kosovichev}, 187--194

\bibitem[{{Field}(1965)}]{Fiel65}
{Field}, G.~B. 1965, \apj, 142, 531

\bibitem[{Goedbloed} {et~al.}(2010){Goedbloed}, {Keppens}, \& {Poedts}]{Goed10}
{Goedbloed}, J. P., {Keppens}, R., \& {Poedts}, S. 2010,
Advanced Magnetohydrodynamics, (Cambridge University Press)

\bibitem[{{Guo} {et~al.}(2010){Guo}, {Schmieder}, {D{\'e}moulin}, {Wiegelmann},
  {Aulanier}, {T{\"o}r{\"o}k}, \& {Bommier}}]{Guo10}
{Guo}, Y., {Schmieder}, B., {D{\'e}moulin}, P., {Wiegelmann}, T., {Aulanier},
  G., {T{\"o}r{\"o}k}, T., \& {Bommier}, V. 2010, \apj, 714, 343

\bibitem[{{Jing} {et~al.}(2010){Jing}, {Yuan}, {Wiegelmann}, {Xu}, {Liu}, \&
  {Wang}}]{Jing10}
{Jing}, J., {Yuan}, Y., {Wiegelmann}, T., {Xu}, Y., {Liu}, R., \& {Wang}, H.
  2010, \apjl, 719, L56

\bibitem[{{Karpen} \& {Antiochos}(2008)}]{Karp08}
{Karpen}, J.~T. \& {Antiochos}, S.~K. 2008, \apj, 676, 658

\bibitem[{{Karpen} {et~al.}(2001){Karpen}, {Antiochos}, {Hohensee}, {Klimchuk},
  \& {MacNeice}}]{Karp01}
{Karpen}, J.~T., {Antiochos}, S.~K., {Hohensee}, M., {Klimchuk}, J.~A., \&
  {MacNeice}, P.~J. 2001, \apjl, 553, L85

\bibitem[{{Karpen} {et~al.}(2006){Karpen}, {Antiochos}, \& {Klimchuk}}]{Karp06}
{Karpen}, J.~T., {Antiochos}, S.~K., \& {Klimchuk}, J.~A. 2006, \apj, 637, 531

\bibitem[{{Karpen} {et~al.}(2003){Karpen}, {Antiochos}, {Klimchuk}, \&
  {MacNeice}}]{Karp03}
{Karpen}, J.~T., {Antiochos}, S.~K., {Klimchuk}, J.~A., \& {MacNeice}, P.~J.
  2003, \apj, 593, 1187

\bibitem[{{Karpen} {et~al.}(2005){Karpen}, {Tanner}, {Antiochos}, \&
  {DeVore}}]{Karp05}
{Karpen}, J.~T., {Tanner}, S.~E.~M., {Antiochos}, S.~K., \& {DeVore}, C.~R.
  2005, \apj, 635, 1319

\bibitem[{{Keppens} {et~al.}(2003){Keppens}, {Nool}, {T{\'o}th}, \&
  {Goedbloed}}]{Kepp03}
{Keppens}, R., {Nool}, M., {T{\'o}th}, G., \& {Goedbloed}, J.~P. 2003, Computer
  Physics Communications, 153, 317

\bibitem[{{Keppens} {et~al.}(2011){Keppens}, {Meliani}, {van Marle}, {Delmont}, {Vlasis}, \& {van der Holst}}]{Kepp11}
{Keppens}, R., {Meliani}, Z., {van Marle}, A.J., Delmont, P., Vlasis, A., \& {van der Holst}, B. 2011, \jcp, in press (doi:10.1016/j.jcp.2011.01.020)

\bibitem[{{Kippenhahn} \& {Schl{\"u}ter}(1957)}]{Kipp57}
{Kippenhahn}, R. \& {Schl{\"u}ter}, A. 1957, Zeitschrift fur Astrophysik, 43,
  36

\bibitem[{{Klimchuk} {et~al.}(2010){Klimchuk}, {Karpen}, \&
  {Antiochos}}]{Klim10}
{Klimchuk}, J.~A., {Karpen}, J.~T., \& {Antiochos}, S.~K. 2010, \apj, 714, 1239

\bibitem[{{Kuperus} \& {Raadu}(1974)}]{Kupe74}
{Kuperus}, M. \& {Raadu}, M.~A. 1974, \aap, 31, 189

\bibitem[{{Lin} {et~al.}(2005){Lin}, {Engvold}, {Rouppe van der Voort}, {Wiik},
  \& {Berger}}]{Lin05}
{Lin}, Y., {Engvold}, O., {Rouppe van der Voort}, L., {Wiik}, J.~E., \&
  {Berger}, T.~E. 2005, \solphys, 226, 239

\bibitem[{{Lin} {et~al.}(2003){Lin}, {Engvold}, \& {Wiik}}]{Lin03}
{Lin}, Y., {Engvold}, O.~R., \& {Wiik}, J.~E. 2003, \solphys, 216, 109

\bibitem[{{Litvinenko} \& {Wheatland}(2005)}]{Litv05}
{Litvinenko}, Y.~E. \& {Wheatland}, M.~S. 2005, \apj, 630, 587

\bibitem[{{L\"ohner}(1987)}]{lohner}
{L\"ohner}, R. 1987, Comp. Meth. in Appl. Mech. and Engineering, 61, 323

\bibitem[{{L{\'o}pez Ariste} {et~al.}(2006){L{\'o}pez Ariste}, {Aulanier},
  {Schmieder}, \& {Sainz Dalda}}]{Lope06}
{L{\'o}pez Ariste}, A., {Aulanier}, G., {Schmieder}, B., \& {Sainz Dalda}, A.
  2006, \aap, 456, 725

\bibitem[{{Mackay}(2005)}]{Mack05}
{Mackay}, D.~H. 2005, in Astronomical Society of the Pacific Conference Series,
  Vol. 346, Large-scale Structures and their Role in Solar Activity, ed.
  {K.~Sankarasubramanian, M.~Penn, \& A.~Pevtsov}, 177--+

\bibitem[{{Mackay} {et~al.}(2010){Mackay}, {Karpen}, {Ballester}, {Schmieder},
  \& {Aulanier}}]{Mack10}
{Mackay}, D.~H., {Karpen}, J.~T., {Ballester}, J.~L., {Schmieder}, B., \&
  {Aulanier}, G. 2010, \ssr, 151, 333

\bibitem[{{Malherbe}(1989)}]{Malh89}
{Malherbe}, J. 1989, in Astrophysics and Space Science Library, Vol. 150,
  Dynamics and Structure of Quiescent Solar Prominences, ed. {E.~R.~Priest},
  115--141

\bibitem[{{Meerson}(1996)}]{Meer96}
{Meerson}, B. 1996, Reviews of Modern Physics, 68, 215

\bibitem[{{Mok} {et~al.}(1990){Mok}, {Drake}, {Schnack}, \& {van
  Hoven}}]{Mok90}
{Mok}, Y., {Drake}, J.~F., {Schnack}, D.~D., \& {van Hoven}, G. 1990, \apj,
  359, 228

\bibitem[{{M{\"u}ller} {et~al.}(2003){M{\"u}ller}, {Hansteen}, \&
  {Peter}}]{Mull03}
{M{\"u}ller}, D.~A.~N., {Hansteen}, V.~H., \& {Peter}, H. 2003, \aap, 411, 605

\bibitem[{{M{\"u}ller} {et~al.}(2004){M{\"u}ller}, {Peter}, \&
  {Hansteen}}]{Mull04}
{M{\"u}ller}, D.~A.~N., {Peter}, H., \& {Hansteen}, V.~H. 2004, \aap, 424, 289

\bibitem[{{Okamoto} {et~al.}(2007){Okamoto}, {Tsuneta}, {Berger}, {Ichimoto},
  {Katsukawa}, {Lites}, {Nagata}, {Shibata}, {Shimizu}, {Shine}, {Suematsu},
  {Tarbell}, \& {Title}}]{Okam07}
{Okamoto}, T.~J., {Tsuneta}, S., {Berger}, T.~E., {Ichimoto}, K., {Katsukawa},
  Y., {Lites}, B.~W., {Nagata}, S., {Shibata}, K., {Shimizu}, T., {Shine},
  R.~A., {Suematsu}, Y., {Tarbell}, T.~D., \& {Title}, A.~M. 2007, Science,
  318, 1577

\bibitem[{{Parker}(1953)}]{Park53}
{Parker}, E.~N. 1953, \apj, 117, 431

\bibitem[{{Patsourakos} {et~al.}(2004){Patsourakos}, {Klimchuk}, \&
  {MacNeice}}]{Pats04}
{Patsourakos}, S., {Klimchuk}, J.~A., \& {MacNeice}, P.~J. 2004, \apj, 603, 322

\bibitem[{{Poland} \& {Mariska}(1986)}]{Pola86}
{Poland}, A.~I. \& {Mariska}, J.~T. 1986, \solphys, 104, 303

\bibitem[{{Priest}(1988)}]{Prie88}
{Priest}, E. R. 1988, Dynamics and Structure of Quiescent Solar Prominences, 
(Kluwer Academic Publishers)

\bibitem[{{Priest} {et~al.}(1996){Priest}, {van Ballegooijen}, \&
  {Mackay}}]{Prie96}
{Priest}, E.~R., {van Ballegooijen}, A.~A., \& {Mackay}, D.~H. 1996, \apj, 460,
  530

\bibitem[{{Schmieder} {et~al.}(2010){Schmieder}, {Chandra}, {Berlicki}, \&
  {Mein}}]{Schm10}
{Schmieder}, B., {Chandra}, R., {Berlicki}, A., \& {Mein}, P. 2010, \aap, 514,
  A68+

\bibitem[{{Schmieder} {et~al.}(1991){Schmieder}, {Raadu}, \& {Wiik}}]{Schm91}
{Schmieder}, B., {Raadu}, M.~A., \& {Wiik}, J.~E. 1991, \aap, 252, 353

\bibitem[{{Serio} {et~al.}(1981){Serio}, {Peres}, {Vaiana}, {Golub}, \&
  {Rosner}}]{Seri81}
{Serio}, S., {Peres}, G., {Vaiana}, G.~S., {Golub}, L., \& {Rosner}, R. 1981,
  \apj, 243, 288

\bibitem[{{Tandberg-Hanssen}(1995)}]{Tand95}
{Tandberg-Hanssen}, E. 1995, The Nature of Solar Prominences, (Kluwer 
Academic Publishers)

\bibitem[{{T\'oth} \& {Odstr\v cil}(1996)}]{toth96}
{T\'oth}, G. \& {Odstr\v cil}, D. 1996, \jcp, 128, 82

\bibitem[{{Townsend}(2009)}]{Town09}
{Townsend}, R.~H.~D. 2009, \apjs, 181, 391

\bibitem[{{van Ballegooijen} \& {Martens}(1990)}]{vanB90}
{van Ballegooijen}, A.~A. \& {Martens}, P.~C.~H. 1990, \apj, 361, 283

\bibitem[{{van der Linden} \& {Goossens}(1991)}]{vand91}
{van der Linden}, R.~A.~M. \& {Goossens}, M. 1991, \solphys, 131, 79

\bibitem[{van Marle} \& {Keppens}(2011)]{vanM11}
{van Marle}, A. J. \& {Keppens}, R. 2011, Computer \& Fluids, 42, 44

\bibitem[{{Vernazza} {et~al.}(1981){Vernazza}, {Avrett}, \& {Loeser}}]{Vern81}
{Vernazza}, J.~E., {Avrett}, E.~H., \& {Loeser}, R. 1981, \apjs, 45, 635

\bibitem[{{Wang} \& {Muglach}(2007)}]{Wang07}
{Wang}, Y. \& {Muglach}, K. 2007, \apj, 666, 1284

\bibitem[{{Withbroe}(1988)}]{With88}
{Withbroe}, G.~L. 1988, \apj, 325, 442

\bibitem[{{Withbroe} \& {Noyes}(1977)}]{With77}
{Withbroe}, G.~L. \& {Noyes}, R.~W. 1977, \araa, 15, 363

\bibitem[{{Wu} {et~al.}(1990){Wu}, {Bao}, {An}, \& {Tandberg-Hanssen}}]{Wu90}
{Wu}, S.~T., {Bao}, J.~J., {An}, C.~H., \& {Tandberg-Hanssen}, E. 1990,
  \solphys, 125, 277

\bibitem[{{Yan} {et~al.}(2001){Yan}, {Deng}, {Karlick{\'y}}, {Fu}, {Wang}, \&
  {Liu}}]{Yan01}
{Yan}, Y., {Deng}, Y., {Karlick{\'y}}, M., {Fu}, Q., {Wang}, S., \& {Liu}, Y.
  2001, \apjl, 551, L115

\bibitem[{{Zirker} {et~al.}(1998){Zirker}, {Engvold}, \& {Martin}}]{Zirk98}
{Zirker}, J.~B., {Engvold}, O., \& {Martin}, S.~F. 1998, \nat, 396, 440

\end{thebibliography}
\end{document}